\documentclass[useAMS,usenatbib]{mn2e}

\usepackage{graphicx}
\usepackage{float}
\usepackage{url}

\title[Sky Quality Meter measurements in a colour changing world
]{Sky Quality Meter measurements in a colour changing world}

\author[A.S\'anchez de Miguel et al.]{A. S\'anchez de Miguel$^{1,2,5,+}$\thanks{E-mail: asanchez@iaa.es}, M. Aub\'e$^{1,+}$\thanks{E-mail: martin.aube@cegeps.pngherbrooke.qc.ca}, J. Zamorano$^{2}$, M. Kocifaj$^{3,4}$, J. Roby$^{1}$, 
\newauthor and  C. Tapia$^{2}$\\
$^{1}$ C\'egep de Sherbrooke, 475 rue du C\'egep, Sherbrooke, J1E 4K1, Canada\\
$^{2}$ Dep. Astrof\'isica y CC. de la Atm\'osfera Universidad Complutense de Madrid, Spain\\
$^{3}$ ICA, Slovak Academy of Sciences, D\'{u}bravsk\'{a} Road 9, 84503 Bratislava, Slovakia\\
$^{4}$ Faculty of Mathematics, Physics, and Informatics, Comenius University, Mlynsk\'a dolina, 84248 Bratislava, Slovakia\\
$^{5}$ Instituto de Astrof\'isica de Andaluc\'ia-CSIC, Glorieta de la Astronom\'ia s/n, 18008 Granada. Spain  \\\\
$^{+}$these authors contributed equally to this work}

\begin{document}

\date{Accepted . Received ; in original form }

\pagerange{\pageref{firstpage}--\pageref{lastpage}} \pubyear{2014}

\maketitle

\label{firstpage}

\begin{abstract}
The Sky Quality Meter (SQM) has become the most common device to track the
evolution of the brightness of the sky from polluted regions to first class astronomical observatories. A vast database of SQM measurements already exists for many
places in the world. Unfortunately, the SQM operates over a wide spectral band and
its spectral response interacts with the sky’s spectrum in a complex manner. This
is why the optical signals are difficult to interpret when the data are recorded in
regions with different sources of artificial light. The brightness of the night sky is
linked in a complex way to ground-based light emissions while taking into account
atmospheric-induced optical distortion as well as spectral transformation from the
underlying ground surfaces. While the spectral modulation of the sky’s radiance has
been recognized, it still remains poorly characterized and quantified. The impact of the
SQM’s spectral characteristics on the sky brightness measurements is here analysed
for different light sources, including low and high pressure
sodium lamps, PC-amber and white LEDs, metal halide, and mercury lamps. We show that a routine conversion of radiance to magnitude is difficult or
rather impossible because the average wavelength depends on actual atmospheric and
environment conditions, the spectrum of the source, and device specific properties.
We correlate SQM readings with both the Johnson astronomical photometry bands
and the human system of visual perception, assuming different lighting technologies.
These findings have direct implications for the processing of SQM data and for its
improvement and/or remediation.
\end{abstract}


\section{Introduction}
The Sky Quality Meter (SQM) is widely used in studies related to light pollution \citep{falchi2011campaign, pun2012night,kyba2012red,kyba2015worldwide,puschnig2014night}. 
Most research has been done over a period of time characterized by a very slow change in illumination technologies and their spectra. With the massive arrival of Light 
Emitting Diodes (LED) in street lighting, the situation is changing very rapidly. We can expect an exponetial growth of the replacements of the incandescent lamps by LED lamps as it happened in the 
70's \citep{riegel1973light}. This projection is confirmed by the ISS observations \citep{firstISS} and the market projections for the indoor illumination \citep{market,katona2016status}. The change of the colour of the sky during the night was detected for the first time by \cite{kyba2012red}. The authors used SQMs with special filters to monitor the change of colour. At that time no link between the colour of the sky and the change in lamp spectra had been established. \cite{sanchezdemiguel2015variacion} outlined a change in the spectra of the sky based on an analysis with an allsky CCD camera with Johnson's filters. To monitor correctly this change, we must have detectors with minimal spectral capabilities, like radiometers with colour filters, colour cameras, or even spectrometers. But such instrumentation is relatively expensive and thus cannot easily be spread all over the world. On the other hand, the SQM is very cheap and easy to use, making it a nice instrument for creating global measuring networks. They are useful to track temporal and geographical variations in Sky Brightness (SB), but without any colour information. In this paper, we want to outline how the broadband spectral response of the SQM reacts to variations in the spectrum of the source for a variety of lighting technologies and for the combination of some other sources of natural origin (e.g. scattered moonlight, background stars, natural airglow). 

In the short term, the spectra of the zenith sky changes slightly with the change in the atmospheric aerosol content but more importantly because of clouds. The extinction of ornamental lights and office windows, car headlights (traffic), the phase of the moon and the angle of its altitude in the sky also play an important role.  The coordinates of the zenith in the galactic coordinate system and the geographical latitude of solar activity also affects the spectrum of the sky.  Generally, the spectra of the street lights do not change significantly in the short term, but on a longer time scale, street lamp conversion/replacement can also modify significantly the spectrum of the sky. Now, the world is experiencing a massive transition in lighting technology, in which entire cities can change from classical high-intensity discharge (HID) lamps (e.g. high pressure sodium or HPS) to white LEDs. Under such important natural and human made changes, how  can we prove that the temporal variations in SB as detected by a broadband instrument like the SQM are representative of the variations in SB that should be detected by the human eye?  Moreover, how can we analyse the human made variation of the SB, excluding the natural variations of the SB? In order to provide basic quantitative tools to improve the tracking of human made or artificial SB, we will analyse the impact of each variable component on the reading of the SQM by simulating the detected signal of the SQM. To accomplish the simulation, we multiply the different source spectra by the spectral response of the SQM-L. Using many analyses, we infer the relationship between the SQM reading with the human eye sensitivity under the scotopic and photopic regimes for different lighting technologies and we compare the relationship of the SQM readings with the Johnson $V$ astronomical photometry band.

\section{The characteristics of the SQM }\label{SQM_char}

The SQM is a simple and portable device intended to facilitate the measurement of the SB by non-highly qualified personnel. Up to now, there have been two generations of this device, and soon a new one will be released. The system is composed of a silicon photodiode as detector (ams-TAOS TSL237S) partly covered by a near-infrared rejection filter. The spectral response of the device tries to mimic that of the human eye under the photopic regime.  The device also has a light-to-frequency converter with a significant response up to the near-infrared. The spectral sensitivity of the photodiode, combined with the transmission of the HOYA CM-500 near-infrared cutoff filter, provides a final spectral response overlapping the Johnson $B$ and $V$ bands used in astronomical photometry (wavelength range from 320 to 720 nm, \cite{johnson1953fundamental}). Although it is very user friendly (aim the photometer to the zenith, push the button, and read the data on the display), it is nevertheless accurate enough to perform scientific research \citep{cinzano2005night}. The SQM photometers have a quoted systematic uncertainty of $10$ per cent (0.1\ mag~arcsec$^{-2}$). Its use has become very popular among researchers and amateur astronomers along with interested members of associations that fight against light pollution.

The SQM-L model, the simplest one, is a hand held photometer intended for gathering data in the field. When used with a photographic tripod, one can measure at different angles and build a map of the all-sky brightness \citep{zamorano2013nixnox}. For mapping a geographical area, one can move to selected locations of a grid, taking the measurements one by one \citep{biggs2012measuring,zamorano2016testing,falchi2016new}. The SQM-LU and SQM-LE units should be linked to a computer by means of USB or ethernet connection, respectively. These models are designed to be used at a fixed monitoring station to continually take and register data. These connected devices could also be used to map extended areas using a vehicle and GPS information.

All the models described above share the same characteristics. The Full Width at Half Maximum (FWHM) of the angular sensitivity is $\sim20^{\circ}$ \citep{cinzano2007report}. The sensitivity to a point source $\sim19^{\circ}$ off-axis is a factor of 10 lower than on-axis. A point source $20^{\circ}$ and $40^{\circ}$ off-axis would register 3.0 and 5.0 astronomical magnitudes fainter, respectively (Unihedron SQM-L manual).

We found a high variability in the shape of the angular response of the SQM. But the FWHM remains constant around $\sim20^{\circ}$. The source of this variability comes from inproper alignment of the optics of the SQM \citep{den2011intercomparisons,den2015stability,bara2015report}.

\begin{figure}
 \includegraphics[width=250px]{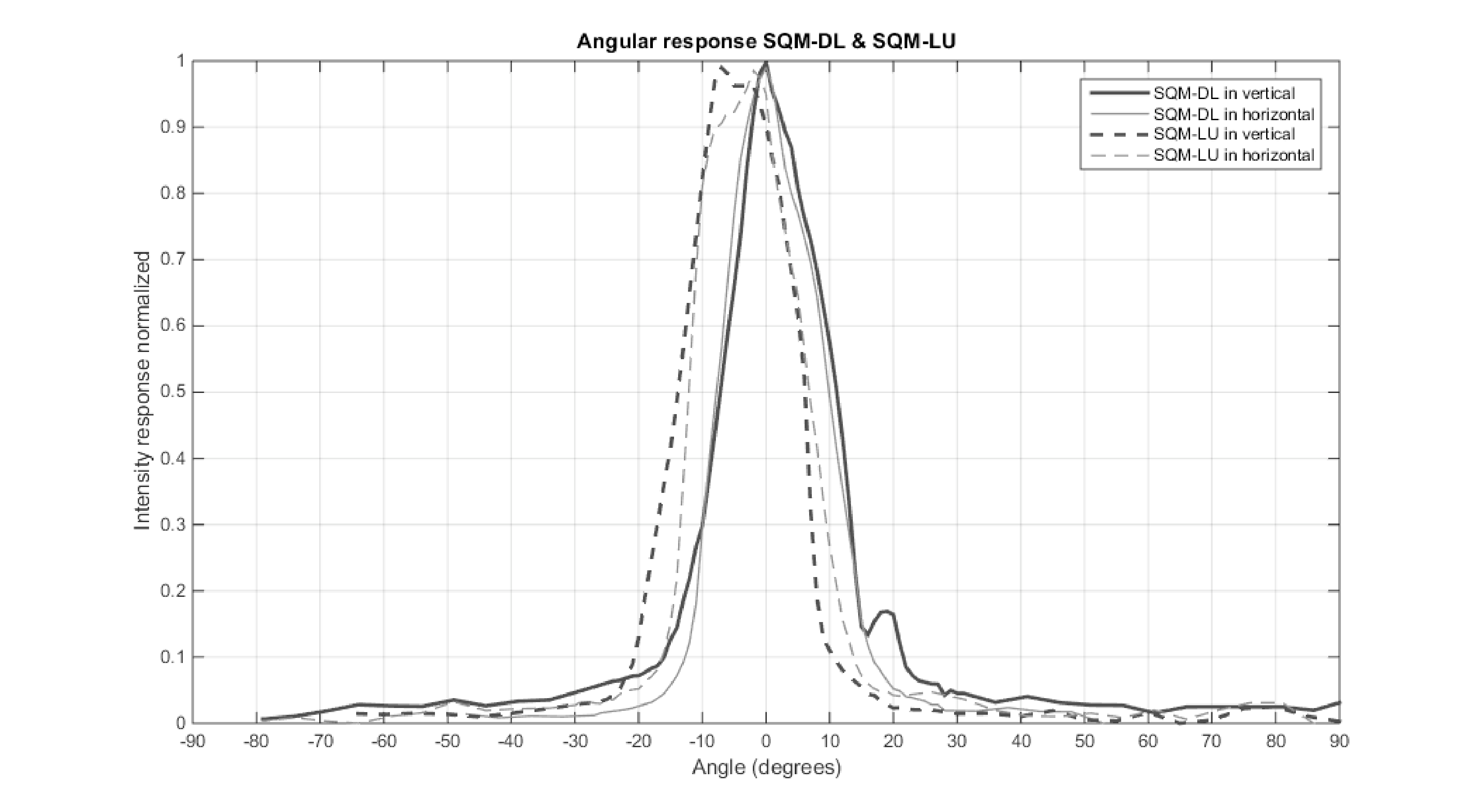}
 \caption{SQM angular response from two SQM in magnitude scale}
 \label{sqmangularresponse1}
\end{figure}

\begin{figure}
 \includegraphics[width=250px]{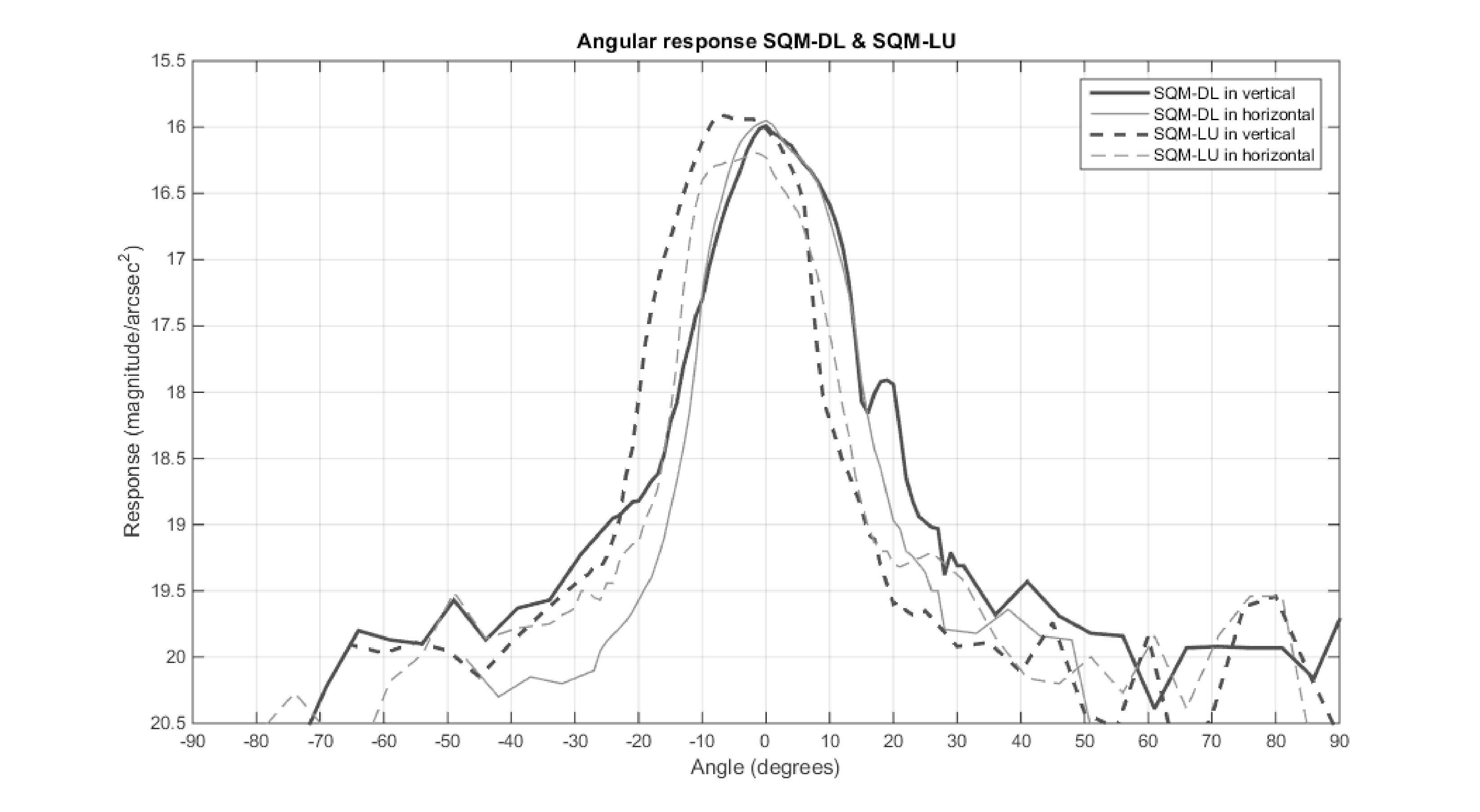}

 \caption{SQM angular response from two SQM in intensity scale}
 \label{sqmangularresponse2}
\end{figure}

The AB magnitudes for any photometric band are determined by Eq.~\ref{ASM1} \citep{fukugita1995galaxy}:

\begin{equation}\label{eq:schechter}
m_{AB}=-2.5 \; \log_{10} \frac{\int_0^\infty T(\lambda)\;\phi(\lambda)\; d\lambda}{\int_0^\infty T(\lambda)\:\phi_{AB}(\lambda)\; d\lambda}
\label{ASM1}
\end{equation}

\noindent where $T(\lambda)$ is the spectral sensitivity of the observation band, $\phi(\lambda)$ is the spectrum of the source and $\phi_{AB}(\lambda)$ is the reference spectrum, defined for a source of constant spectral density flux of 3631 Janskys over the entire spectral range of the band.

The conversion from AB magnitudes \citep{oke1983secondary}, which is an astronomical magnitude system, to radiance ($R$) can be done using the following relationship (this work):

\begin{equation}\label{eq:AS1}
m_{AB}=-2.5 \; \log(R) -5 \log\bar{\lambda}-2.41
\end{equation}

\noindent where $R$ is expressed in 
erg\ s$^{-1}$\ cm$^{-2}\mbox{\AA}^{-1}$
, and $\bar\lambda$ is the average wavelength of the band. The average wavelength is defined by the following equation.

\begin{equation}
\label{averagew}
\bar{\lambda}=\frac{\int_0^\infty T(\lambda) \lambda d\lambda} {\int_0^\infty T(\lambda) d\lambda}
\end{equation}
where $T(\lambda)$ is the spectral sensitivity of the observation band (including the detector response).

Using Eqs \ref{ASM1} and \ref{eq:AS1} we can create synthetic photometry measurements from any spectral source and any spectral band.

The SQM calibration is made using the manufacturer reference lamp\footnote{Anthony Tekatch, Unihedron, personal communication.}. To convert the SQM magnitudes to astronomical $V$ Johnson magnitudes, we need to apply a shift of 0.35 magnitudes \citep{cinzano2005night}. However, \citep{cinzano2005night} has used the Vega magnitudes, which produces problems for converting into the SI units, so we will use the AB magnitude system, which does not have that problem. So we only need to use the transformation $(SQM-V)_{AB}=(SQM-V)_{VEGA}-0.04$ based on Eq.~\ref{averagew}.  Hence, the SQM offset between the $(SQM_O-V)_{VEGA}=(SQM-V)_{VEGA}-0.35=(SQM-V)_{AB}-0.31$, where $SQM_O$ are the $mag_{SQM}$ from the readings of the SQM devices. This should be considered as a first order correction because a correct relationship needs to take into consideration the spectra of the source. In this paper, we will write $SQM_{AB}$ to express the SQM reading using the AB magnitude definition, and $SQM_O$ to express the SQM readings with the instrument factory calibration as done by Unihedron. Knowing this, we can convert easily from $SQM_O$ to radiance units with Eq.~\ref{eq:AS1} by

\begin{equation}\label{eq:ASX}
SQM_O+0.31=SQM_{AB}=-2.5 \; \log(R) -5 \log\bar{\; 5418\AA\; }-2.41
\end{equation}

From this equation is possible to get, through basic operations,

\begin{equation}\label{eq:ASX2}
R'=0.0270038\cdot10^{-0.4\; SQM_O}
\end{equation}

\noindent where $R$ is in units of erg s$^{-1}$\ cm$^{-2} \AA^{-1}$~arcsec$^{-2}$, $R'$ is in  units of W cm$^{-2}$\ sr$^{-1}$, $SQM_O$ is in units of $mag_{SQM}$~arcsec$^{-2}$, and $5418\AA$ is the effective wavelenght of the SQM band.

\begin{figure}
 \includegraphics[width=250px]{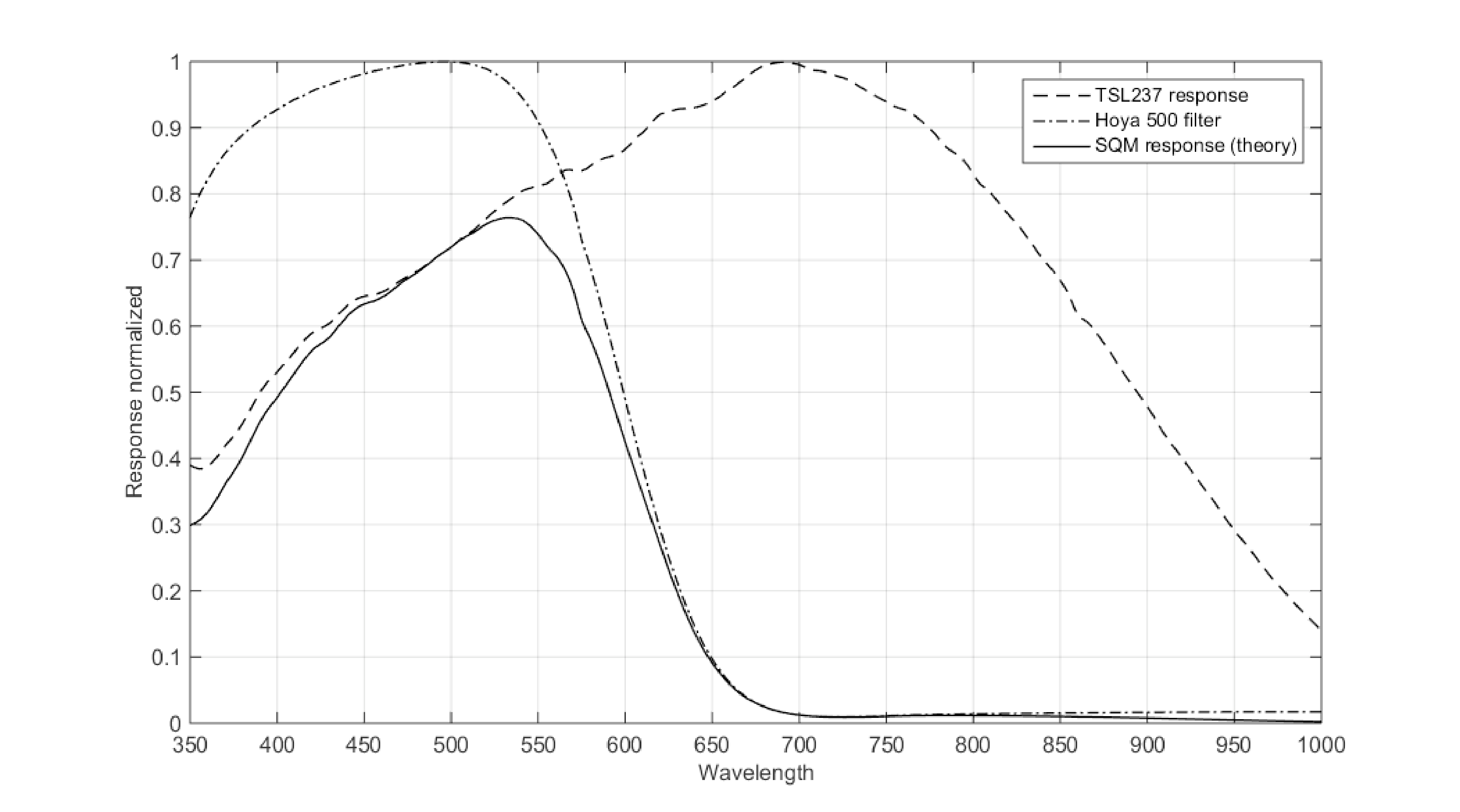}
 \caption{Spectral response of some optics components of the SQM }
 \label{sqmresponse1}
\end{figure}

\begin{figure}
 \includegraphics[width=250px]{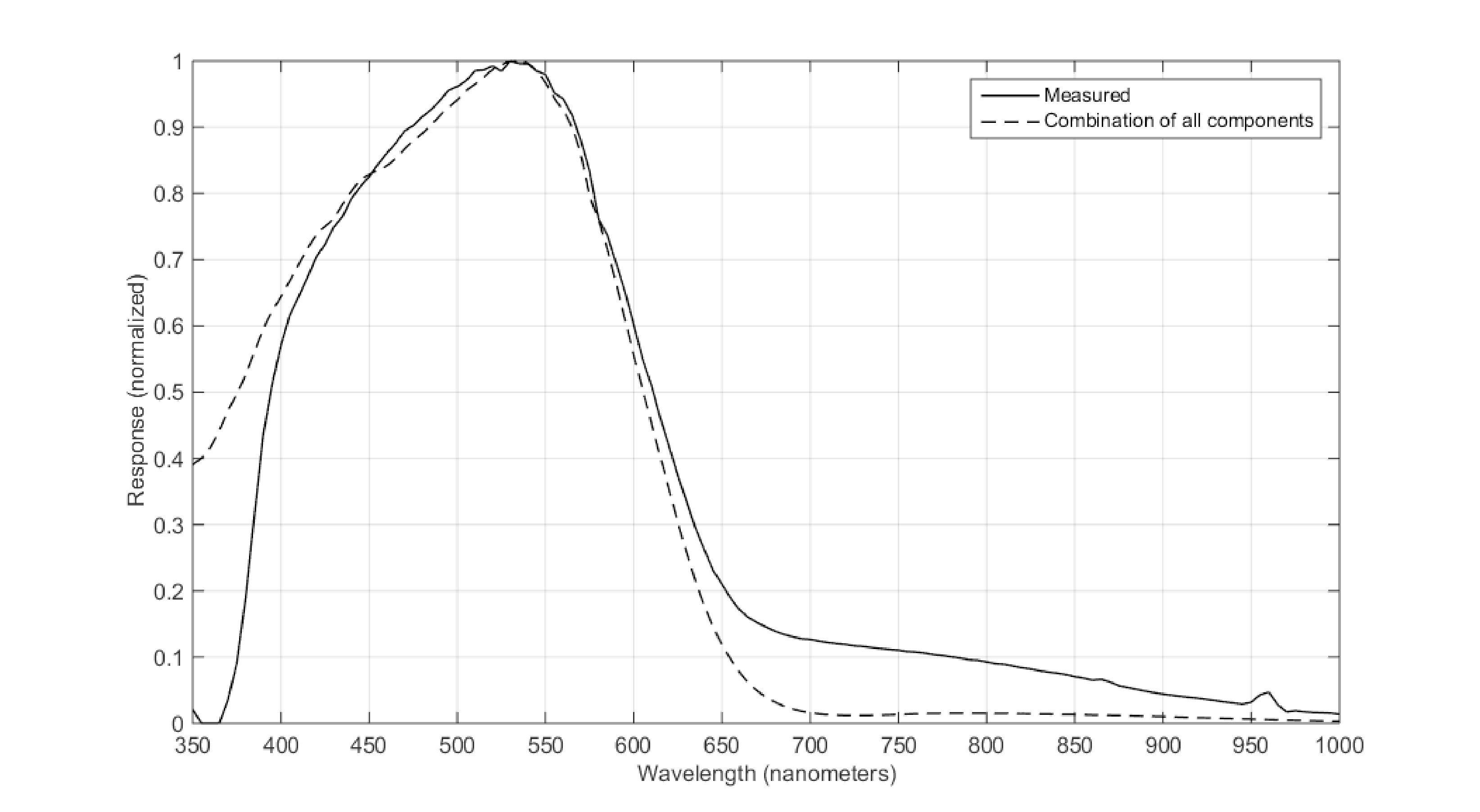}
 \caption{SQM Spectral response. The dashed line correspond to the combination of all the spectral response/transmision of the components while the black line is the total measured response.}
 \label{sqmresponse2}
\end{figure}

We have combined the spectral tranfer functions of the optical parts of the SQM photometers to build the theoretical or expected response of the device.  Instead of using the manufacturer's information, we have measured the spectral response of the detector and the transmission curve of the filter. The window, the infrared rejection filter (Hoya 500) and lens assembly of one SQM photometer were removed to determine the spectral response of the TSL237 detector itself (see Fig.~\ref{sqmresponse1}). The light beam from a monochromator was fed to an integrating sphere with the SQM detector observing through one other port of the sphere. 
\\
\\
The measured SQM total response should be the same as the spectral response that we have determined by combining the measured responses for each optical components. Our experiment shows that important differences arise in the form of a prominent infrared tail (see Fig.~\ref{sqmresponse2}) in the total SQM response. We had similar results for three SQM units that we had in hand (SQM-LU, SQM-LE and SQM-DL). This unexpected near infrared response is caused by the fact that the infrared rejection filter does not fully cover the optical path. The Hoya 500 filter can be recognized as a square blue glass located behind the lens assembly. Also there are a lack of sensitivity on the UV parte, probably because of the transmitance of the lens used.
\\

\subsection{SQM readings vs the photopic, scotopic vision, and Johnson photometric systems}

\begin{figure}
 \includegraphics[width=250px]{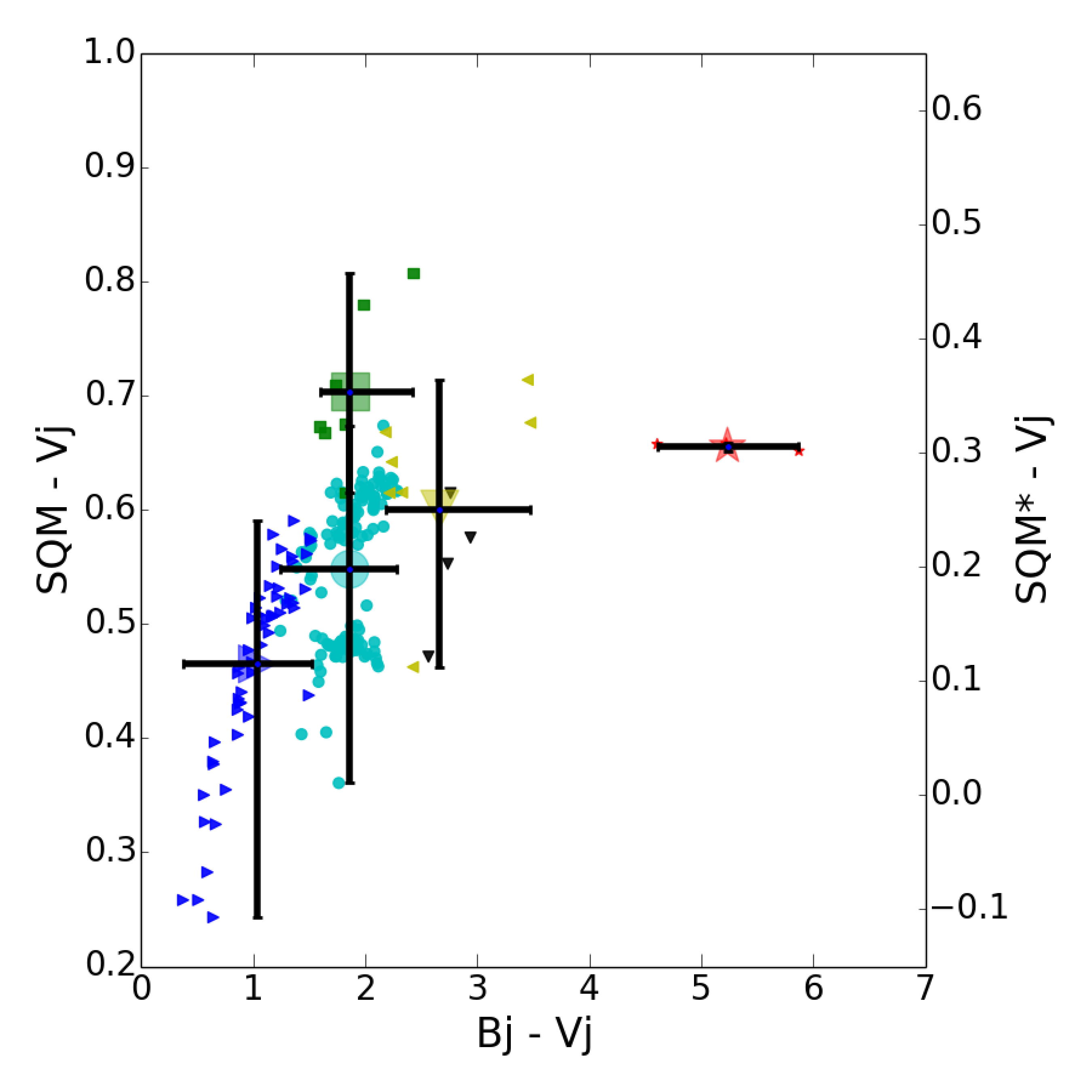}
 \caption{Colour--colour diagram of the SQM minus $V_j$ Johnson band versus the Johnson $B_j$ minus $V_j$ colour index for different spectra of typical street lights. The right scale of the figure is offset by 0.35 magnitudes from the left scale because of a hardware offset associated with the SQM  \citep{cinzano2005night}. The \textbf{red} stars correspond to low pressure sodium lamps, pure amber LED and others; the \textbf{black} down pointing triangles are standard high pressure sodium lamps; the \textbf{yellow} left pointing triangles are high pressure lamps without Hg, high pressure lamps with phosphor,  very warm ceramic metal halide and PC-amber lamps(CCT $<2200$ k); the \textbf{green} squares correspond to mercury lamps, \textbf{cyan} circles warm white LED(CCT 2200 - 3500 k), and normal warm ceramic metal halide; the \textbf{blue} right pointing triangles correspond to cool white LED(CCT 3500 - 9000 k), standard metal halide and ceramic metal halide. The error bars correspond to the most extreme values of each categories.
 \label{color1}
}
\end{figure}

Ever since \cite{johnson1953fundamental} defined the first astronomical photometric system, the sky brightness has been measured by astronomers. Magnitudes in the AB system can be converted to other systems. However, because all magnitude systems involve integration of some assumed source spectrum over some assumed spectral range or band, such conversions are not necessarily trivial to determine. Various authors have computed conversions for standard situations.  \cite{johnson1953fundamental} clearly explained that even if the colour of the $V_j$ filter is yellow, it does not mean that the $V_j$ is representative of the human vision response. In fact, the $V_j$ band peaks at the central point of the daylight human vision but the bandwidth is not broad enough to represent all the contribution to the human vision. The accepted daylight human light curve, dominated by the cones, was set by CIE 1931 \citep{smith1931cie} as the $V(\lambda)$ curve, also known as photopic vision. The low light level vision, dominated by the rods, has a different response and is called scotopic vision \citep{judd1951report}. Now the CIE has recently approved an intermediate vision regime based on \cite{rea2004proposed} and \cite{eloholma2006new}, the so-called mesopic vision. Mesopic vision involves both cones and rods with various proportions depending on the light level.

In order to show the systematic differences in this secction we are going to compare the diferent results that the SQM gives when it recibes direct light from a sample street lights, and we are going to compare this results with the real problem that the SQM would ideally to solve that is the stars visivility by humans. 

In Fig.~\ref{color1}, we can see how  the $SQM_{AB}-V_j$ depends on the $B_j$--$V_j$ colour. Excluding the effect of the natural spectrum of the sky ((i.e. the case that the sky brightness is dominated by the light pollution), ignoring the lamp spectra results in a span of values as large as 0.55 magnitudes along the $SQM_{AB}-V_j$ axis ($\sim 0.25$ to $\sim 0.8$). Even if we know the technology of the lamp (e.g. the same symbol in Fig.~\ref{color1}), we observe a large span of $SQM_{AB}-V_j$ values for the same technology class. The span of $SQM_{AB}-V_j$ values can increase up to 0.35 magnitudes for the lamps that include more blue light (blue right pointing triangles in Fig.~\ref{color1}), 0.32 for the whitish lamps (cyan circles in Fig.~\ref{color1}), and even 0.25 magnitudes for the yellowish lamps (yellow left pointing triangles in Fig.~\ref{color1}). Almost all the lamps are concentrated in the region $SQM_{AB}-V_j = 0.55\pm0.20$.  

\begin{figure}
 \includegraphics[width=250px]{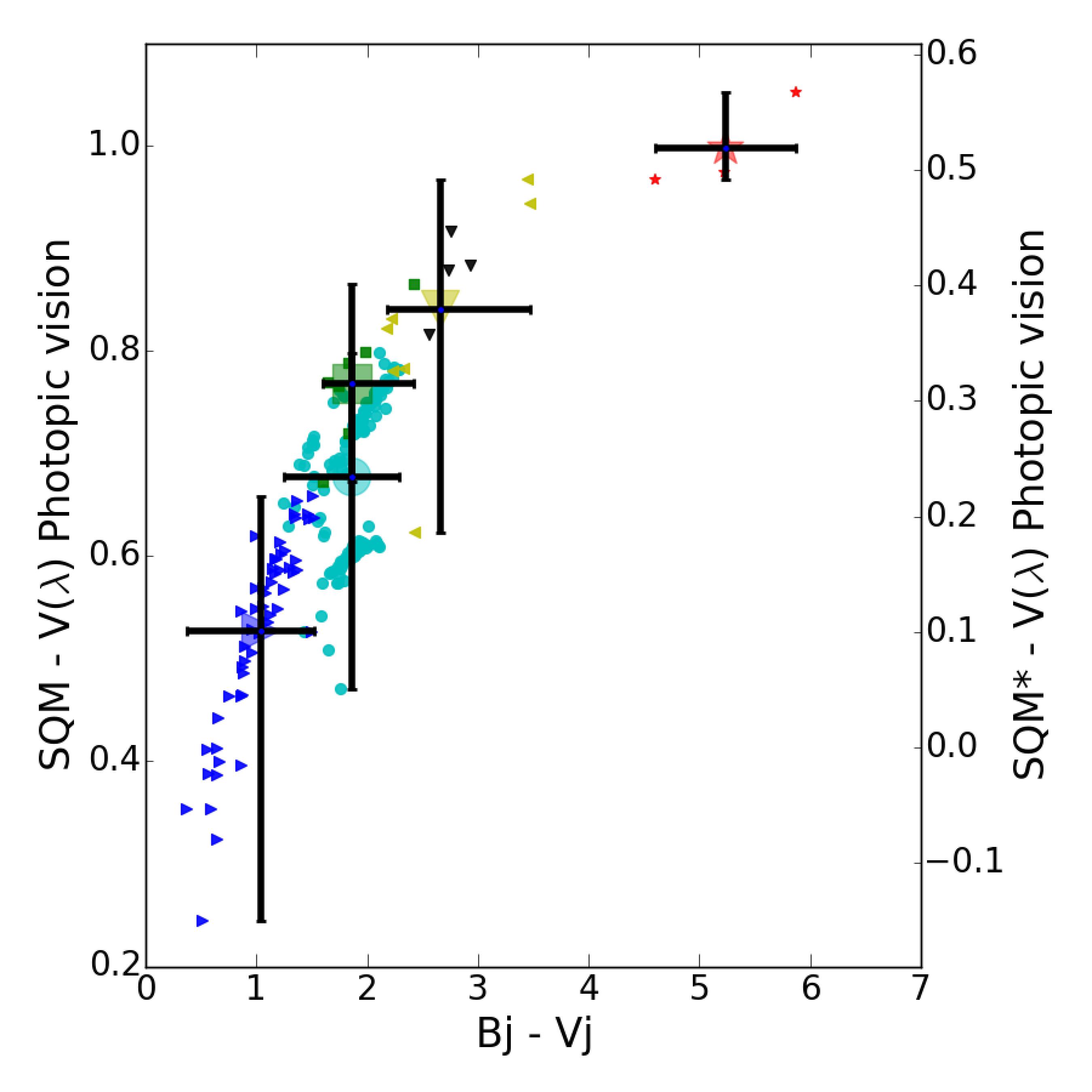}
 \caption{Colour--colour diagram of the SQM minus $V(\lambda)$ the CIE 1931 photopic curve versus the colour Johnson $B_j$ minus $V_j$ for different spectra of typical street lights. Same colour code and symbols as Fig.~\ref{color1}.}
 \label{color2}
\end{figure}

It is possible to see that there is an important difference between Fig.~\ref{color1} and Fig.~\ref{color2}. The latter shows the $SQM_{AB}-V(\lambda)$ colour diagram. Indeed, the dispersion of the values along the vertical axis is larger when expressed with the $V(\lambda)$ curve compared to the use of the $V$ Johnson band (0.72 vs 0.55). It is also interesting to see that for the category of white lamps (cyan circles), the lamp population appears to be split into two sub-classes in both figures. Another striking feature of the plots is the important differences in the vertical axis values between each technology/colour. There is also a systematic offset in the vertical axis values between the two plots. For the greenish lamps we have an offset of 0.05 while it is 0.15 with the whitish lamps, the higher values obtained in Fig.~\ref{color2}. 

The SQM has a spectral sensitivity that is in between the $B_j$ and the $V_j$ bands. It is assumed by the manufacturer that it is a better tracer of the scotopic vision than the $V_j$ band. It is noticeable from Fig.~\ref{color2} that if one does not know the kind of lamp/colour, an error as large as 1.2 mag can be made when using SQM to estimate the scotopic magnitude $V(\lambda)$. With the SQM, the sky appears brighter for the reddish lamps in comparison to the photopic perception.

\begin{figure}
 \includegraphics[width=250px]{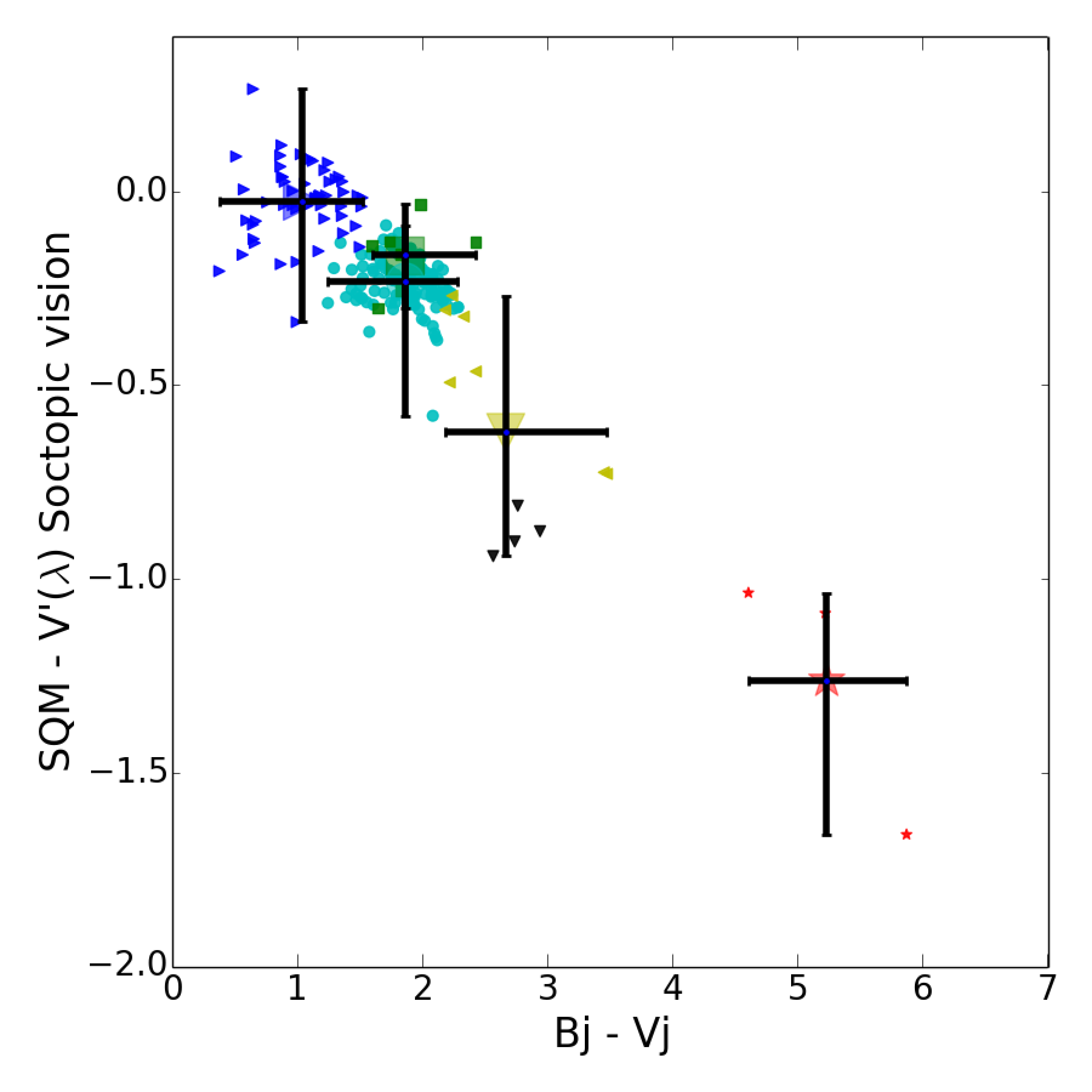}
 \caption{Colour--colour diagram of the SQM minus $V’(\lambda)$ CIE 1951 scotopic vision versus the colour index $B_j$ minus $V_j$ for different spectra of typical street lights. In that case there is no offset between the calculated and instrumental magnitudes. Same colour code and symbols as in Fig.~\ref{color1}. \label{color3}
}
\end{figure}

Fig. \ref{color3}, which gives the SQM reading in reference to the scotopic system, shows that the relationship goes in the opposite direction than in Fig.~\ref{color2}. The figure shows that if we use the SQM to evaluate the perception in the scotopic regime, the sky appears darker for the reddish lamps. The large dispersion of the sodium-like lamps is noticeable in Fig.~\ref{color3}, with more than 0.2 magnitudes of dispersion. Figs \ref{color2} and \ref{color3} show clear correlations with the $B_j-V_j$ colour index. This means that it is more accurate to measure the colour to estimate the $SQM-V(\lambda$) or $SQM-V'(\lambda$) than just guess it from a technology based averaged value. In other words, if a correction is going to be made to the SQM readings, to consider the tecnology of the lamp is not acurate enough, but, using the $B_j-V_j$ colour, is posible to correct properly the measurements of an SQM. So, if using a color sensitive instrument like a DSLR is posible to get an estimation of the tipcal $B_j-V_j$ colour of the light pollution of the place, an accurate correction can be applied, although a lot of scatter will remain.

On the light polluted site of the Universidad Complutense de Madrid (UCM) observatory (40$^o$ 27' 04'' N, 03$^o$ 43' 34'' W), where we can have an Sky brightness measurement on clear conditions of $SQM_O$ = 18, the light pollution was originally dominated by HPS lamps, so that the photopic magnitude was 17.15 (18-0.85)\footnote{these values should be read as: sky brightness on the band$(SQM_{O}\pm$ correction)}, but the scotopic magnitude was 18.6 (18+0.6). Since then, the street lights have been gradually changed to blueish lights, but the SQM measurements have still maintained the same readings. The new photopic magnitude assuming pure blueish lamps should be 17.5 (18-0.5), so it should look darker than before, but with adapted vision we get a scotopic magnitude of 18 (18+0), so it should look brighter. We need to be aware of this effect when we are trying to track the final visual perception of the sky brightness.

\subsubsection{Simulation of SQM readings of a realistic sky spectrum}

In the previous subsection we have shown how the SQM readings can be compared with the direct incident light from lamps. A realistic simulation of the SQM response to a variety of spectrum needs to take into consideration several factors, for example the contribution of the natural sky, the Rayleigh and Mie scattering by air molecules and aerosols. Therefore, to simulate a more realistic scenario we have added the scattered light from different lamp spectrum at various distances (3 km\footnote{City edge.}, 6 km , 15 km, 30 km, 60 km and 100 km). Also, we applied the Kocifaj model(MSNsRAU \cite{aube2012using}) using the constants from  \cite{aube2013evaluating} and combined this with the natural component spectra from the Montsec Observatory acquired with the SAND spectrograph. 

\begin{figure}
 \includegraphics[width=250px]{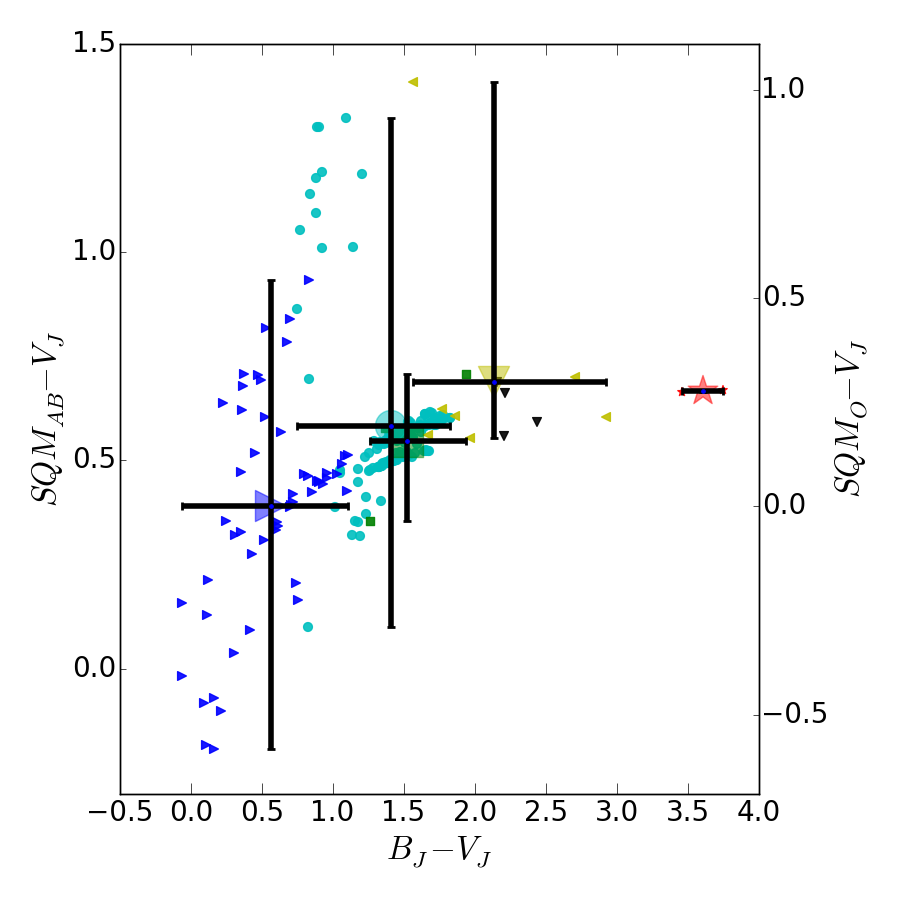}
 \caption{Colour-colour diagram of the SQM minus $V_j$ Johnson band versus the Johnson $B_j$ minus $V_j$ colour index for different spectra of typical street lights scattered by the atmospher at 3 kilometers. The right scale of the figure is offset by 0.35 magnitudes from the left scale because of a hardware offset associated with the SQM  \citep{cinzano2005night}. The \textbf{red} stars correspond to low pressure sodium lamps, pure amber LED and others; the \textbf{black} down pointing triangles are standard high pressure sodium lamps; the \textbf{yellow} left pointing triangles are high pressure lamps without Hg, high pressure lamps with phosphor,  very warm ceramic metal halide and PC-amber lamps; the \textbf{green} squares correspond to mercury lamps, \textbf{cyan} circles warm white LED, and normal warm ceramic metal halide; the \textbf{blue} right pointing triangles correspond to cool white LED, standard metal halide and ceramic metal halide. The error bars correspond to the most extreme values of each categories.}
 \label{colorSQMVj3k}
\end{figure}

In both, Fig. \ref{colorSQMVj3k} and Fig. \ref{color1} a comparison is made between the SQM and V Johnson colour and the B Johnson and V Johnson colour of the streetlamp spectrum. Figure \ref{colorSQMVj3k} concerns scattered light produced by a large city at a distance of 3 kilometers and Figure \ref{color1}  concerns direct light. The data shows clearly that when light sources have a redder colour the scattering does not cause a significant chance between the SQM band readings and the V Johnson band values. In the same two figures it becomes evident that the bluer coloured lamps cause two significant changes. Firstly, lamps with a significant infrared emission have a completely different behaviour than the pure visible light emission due to the low atmospheric scattering in the infrared. Secondly, in figure \ref{color1} the colour difference between the redder and bluer lamps is $\sim0.5$ magnitudes. In figure 8 this has increased to $\sim0.7$ magnitudes. What is important to mention is that in figure \ref{colorSQMVj3k} the extremer values up to 1.5 magnitudes are not caused by the redder or bluer lights but by the lamps with more infrared emission as well as the bluer lamps without infrared emission.

\begin{figure}
 \includegraphics[width=250px]{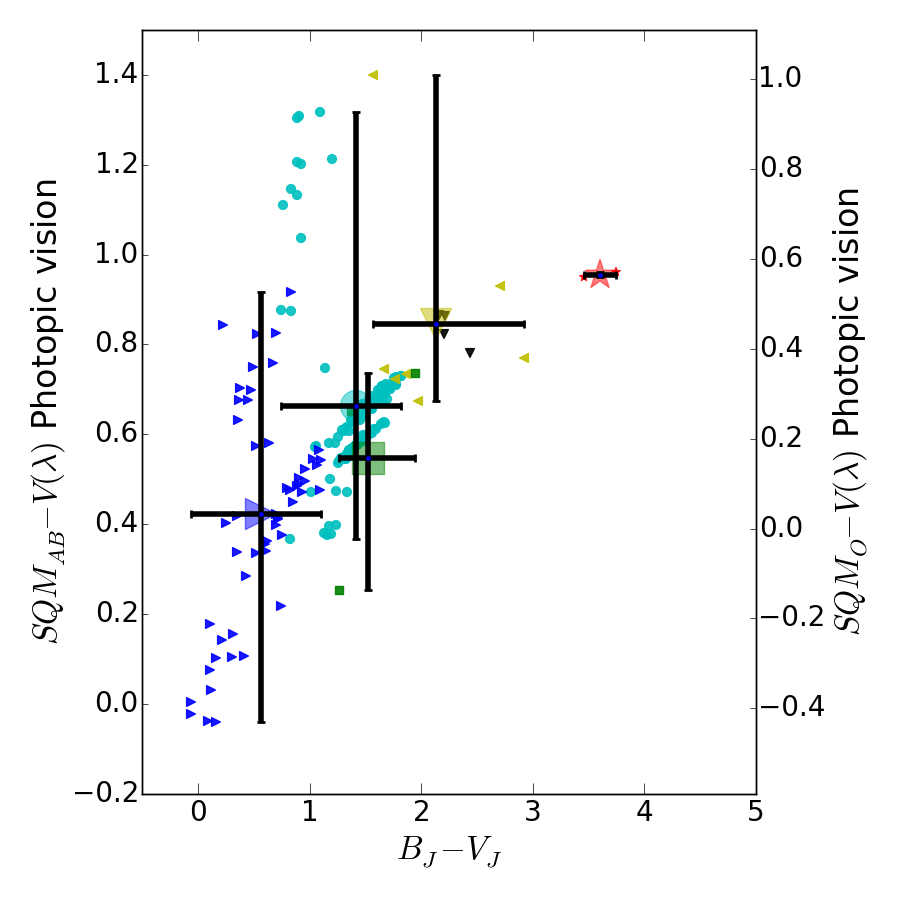}
 \caption{Colour--colour diagram of the SQM minus $V(\lambda)$ the CIE 1931 photopic curve versus the colour Johnson $B_j$ minus $V_j$ for different spectra of sky polluted by at 3 km of the source by typical street lights. Same colour code and symbols as Fig. \ref{color1}.}
 \label{colorSQMVl3k}
\end{figure}

\begin{figure}
 \includegraphics[width=250px]{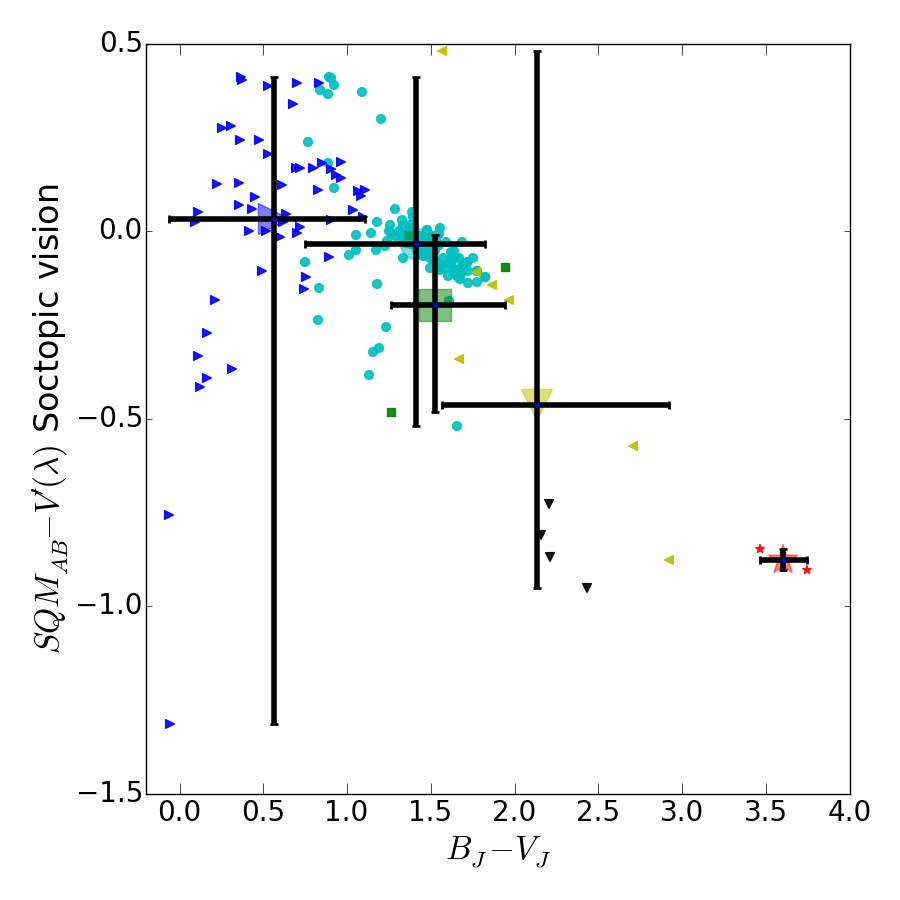}
\caption{Colour--colour diagram of the SQM minus $V’(\lambda)$ CIE 1951 scotopic vision versus the colour index $B_j$ minus $V_j$ for different spectra of sky polluted by at 3 km of the source by typical street lights(city edge). In that case there is no offset between the calculated and instrumental magnitudes. Same colour code and symbols as in Fig.~\ref{color1}.
}
 \label{colorSQMSC3k}
\end{figure}

The comparison between figure \ref{color2} and \ref{colorSQMVl3k} is a similar case as the comparison between Fig. \ref{colorSQMVj3k} and Fig. \ref{color1}. 
Figure 10 depicts a different situation as the data of SQM readings against the scotopic magnitudes does not diverge from one another. 
Although, the data inside figure \ref{colorSQMSC3k} is more dispersed, it still leads to the same conclusion as figure \ref{color3}. Mainly that the lamps with bluer light cause more sky brightness than lamps with redder light. It is important to highlight that, unlike lamps with yellower lights, all lamps with white lights, including LED and regardless of their technology, systematically show to create brighter skies with 0.4 magnitudes according to the SQM readings. Keep in mind that SQM readings vary from actual scotopic vision, even if for cool lamps produce the right values, because the SQM is unable to detect variations in colour and this may lead to misinterpretations. 

\section{Effect of diferent technologies of ilumination on the skybrightness and its impact on SQM monitoring}

\begin{figure}
 \includegraphics[width=250px]{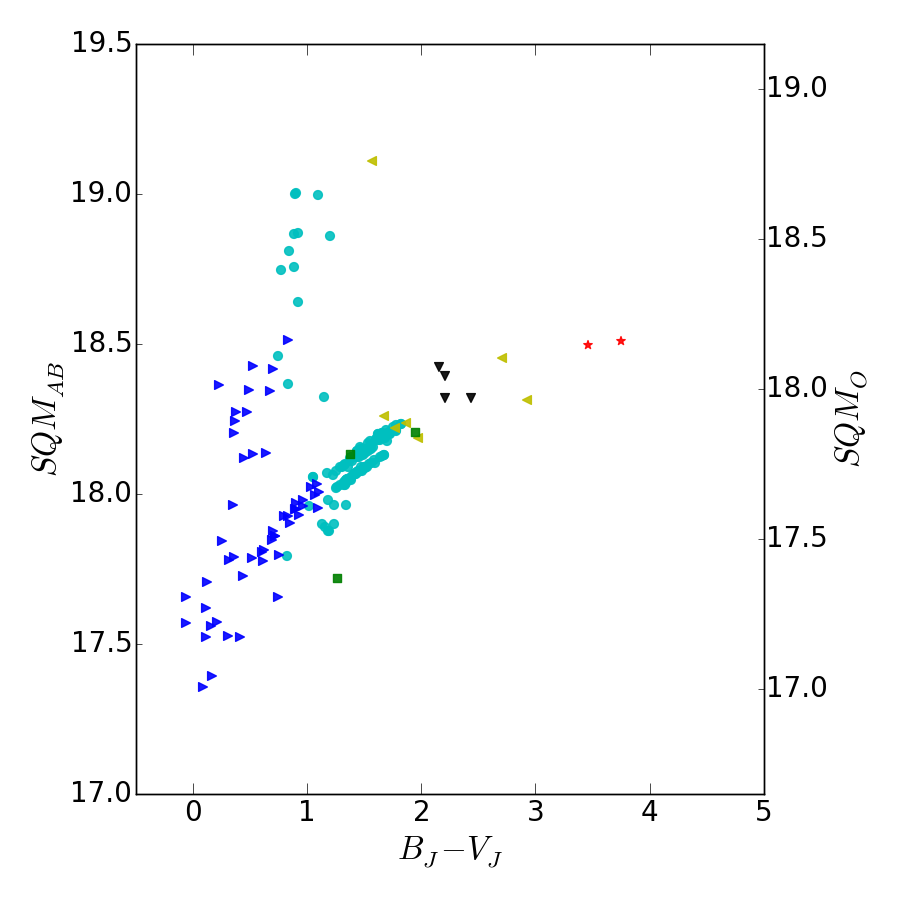}
 \caption{SQM readings simulation from a realistic polluted sky by at 3 km of the source by typical street lights. Same colour code and symbols as Fig.~\ref{color1}.}
 \label{colorSQM3k}
\end{figure}

As we explained in the introduction, we are living during a time of revolution when it comes to illumination technology. With synthetic photometry we are able to simulate how the scattering and spectrum of new lighting systems impact sky brightness. Similar to our method in the previous section, we created a realistic simulation of the sky spectrum. In the experiment that we will be discussing now, we have normalised the spectrums to the same lumens emission before applying the scattering process and adding the natural component. In Figure \ref{colorSQM3k}  it shows that for the same lumens, the bluer lamps create much brighter SQM readings, but even so, we have to keep in mind that the SQM is not  giving us the right Scotopic magnitudes. Therefore, if we measure SB of $SQM_O$=18.2 in a high pressure sodium dominant place at 3 Kilometers distance of the main source, the real scotopic brightness is 19 according to Figure \ref{colorSQMVl3k}. If the illumination is changed suddenly to a bluer LED, our SQM will measure $SQM_O$=17 to 17.5 and the scotopic magnitude will remain the same. This means that according to the SQM readings we got a brighter sky by 0.7 to 1.2 magnitudes, but in reality we got a brighter sky by 1.5 to 2 magnitudes on the human eyes when we change from orange light to white light with content of blue light.

\begin{figure}
 \includegraphics[width=250px]{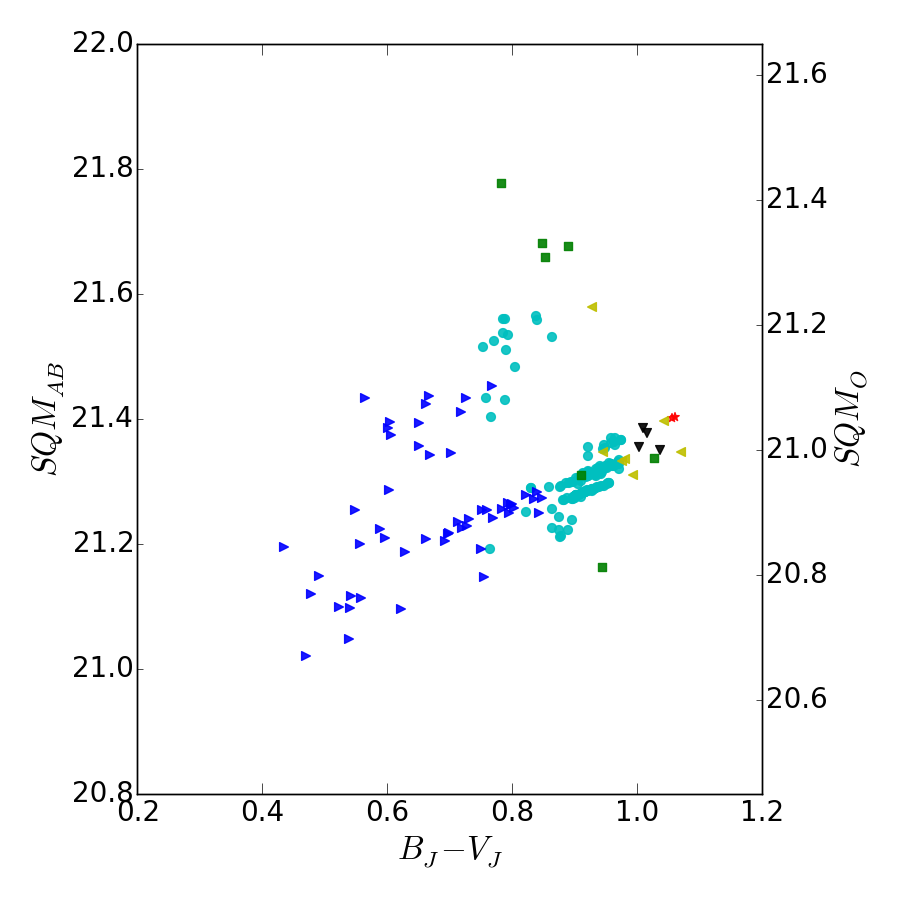}
 \caption{SQM readings simulation from a realistic polluted sky by at 60 km of the source by typical street lights. Same colour code and symbols as Fig.~\ref{color1}.}
 \label{colorSQM60k}
\end{figure}

This effect is less pronounced when we go further away from the source, the absorption of the blue light, makes that the yellowish light can go further. This explains why in Figure \ref{colorSQM60k} the SQM readings of 0.4 magnitudes are lower than before. The new illumination technologies are impacting the Zenith and this affects the SQM readings. 
When we compare old illumination technologies with new illumination technologies and combine both with a fixed amount of light emitted into the atmosphere as well as fixed illumination levels we were able to conclude that the newer illumination technology with more blue content creates brighter skies. Even in the case on \cite{kollath2016qualifying}, where the previous street lighting had big ULOR and no blocking effect, the new ilumination compensate the better photometry with the more contaminant effect of the blue light.

This example shows how, even in places, far removed from the polluting source, the effect of the change of the illumination is still significant when it concerns the SQM readings. This visible effect is easier to observe when the light pollution is stronger. The real effects of the two methods currently used to reduce both the light emission into the upper hemisphere and the illumination levels of new LED lighting systems cannot be fully simulated in this article because of its complexity. The main goal of this article is to create awareness with SQM users and to inform them that when the SQM readings are lower due to less light emitted upwards or lower illumination levels, this does not necessarily mean that the Sky brightness quality has improved. In reality, the sky brightness quality may have even deteriorated. It is therefore vital that the effect of the SQM spectral response, as explained in this article, should be considered before interpreting low SQM readings. To help with future diagnostics of SQM data we have created an extra sample of simulations that can be found as additional information to this article.

\subsection{Aplication to a practical case: Madrid and Milan}
In order to reduce energy consumption and light pollution the city of Madrid has recently replaced most of its midsize high pressure sodium street lamps with LED 3000k lamps. These midsize street lamps account for 40\% of the total amount of street lamps in the city. There is a current plan to replace the remaining 60\% of street lamps sometime in the near future. Meanwhile, the city of Milan has recently replaced their old street lights with LED lamps of 4000k. But are these new lamps really reducing light pollution as has been claimed?

In the table \ref{my-label} it is evident that the new replacement lamps increase the sky brightness for the adapted human vision by 1.04 magnitudes in the case of the 3000k lamps, and by 1.25 magnitudes in the case of 4000k lamps. In a linear scale, the sky brightness for the LED 3000k increases by 161\% and for the LED 4000k by 216\%.

The current retrofits both in Madrid as well as in Milan are solely focusing on replacing the lamp and not the full illumination system. The most typical street luminaires used on the roads of Madrid is the cobra head with 5\% ULOR (Upper Ligh Output Ratio), but there are alternative luminaires available that emit 0\% ULOR. The next modelisation consider the light emitted upwards and the horizontal emission that 
is blocked by surrounding buildings. It is not taking into account any other objects that may block emission such as trees. 
Also, in this section we will refer to the real sky brightness as the sky seen by a human with scotopic vision, because it is the kind of vision that the humans really use to observe the sky.

When we change not only the lamps but also the luminaire with a type that emits 0\% ULOR, the light emission towards the street will automatically increase. As a result the energy output can be decreased to compensate for this emission increase, thus saving on energy consumption. And, if the albedo value of the ground was 0.2 for example, the new total emission towards the sky would be just 80\% of the current emission.

If the city of Madrid was to change the current luminaires. The new luminaries have 0\% ULOR, combined with the new LED 3000k lamps. Would this reduce light pollution? And how accurately would Madrid´s SQM station be able to measure these changes? As was discussed earlier in this article, the SQM is unable to track the real adapted human vision and this means that the SQM readings are not always reflecting the reality. For example, in the case of the 3000k lamp the SQM will indicate no increase of sky brightness (SQM band) when in reality there is an increase of 111\%, and in the case of the 4000k lamps, the SQM will indicate an increase of 14\% (SQM band) when in reality it is 153\%.

 In the case of Madrid, the emission of the LEDs can be easily adjusted with a regulation system. If the possibility to regulate the emission down to 60\% of the original illumination level is combined with the use of luminaires of 0\% ULOR, the 3000k lamps would appear to have a 45\% reduction according to the SQM, but in reality there is an emission increase of 29\%. In the case of the 4000k lamps, the SQM would show a reduction of 30\% when in reality there is an emission increase by 56\%. 

We can conclude from the previously discussed measurements that light pollution has increased due to the new LED lamps. With the current numbers, it is safe to say that the 3000k and 4000k lamps would only reduce light pollution if they were to be switched off or dimmed down to an extremely low intensity. To be more specific, if changing HPS lamps to LED 3000k lamps is meant to reduce light pollution it would be necessary to reduce street illumination to below 42\% of the original level. Such low illumination levels could cause in certain streets and places to become so dark that citizens no longer experience a sense of safety 
and security. This happened in Madrid according to the Madrid city council. That is why 60\% is the dimmest level used in Madrid nowadays. 
This claim is not shared by scientific community but nevertheless is commonly used by technicians.

\begin{table*}
\centering
 \begin{minipage}{160mm}
\caption{This table sumarize the results of the predicted sky brightess on three observation band, the photopic, scotopic and SQM. This simulation is consider for the skybrightness mesured at 3 km of the center of a city like Madrid(city edge). Also is included a reference the change of signal that each band will recibe on linear scale relative to the sodium high presure lamp with 5\%. On the last four cases the light emission levels (amount of lumens needed to produce a given ilumination level) has been reduced to try to 
find the break even point where the blue light effect is compensated by the reduction of the ilumination level.}
\label{my-label}
\begin{tabular}{|l|cccccc}
\hline
Lamp                           & \multicolumn{1}{l|}{SQM(Mag)} & \multicolumn{1}{l|}{SQM} & \multicolumn{1}{l|}{Photo(Mag)} & \multicolumn{1}{l|}{Photo} & \multicolumn{1}{l|}{Scoto(Mag)} & \multicolumn{1}{l|}{Scoto} \\ \hline
HPS ULOR 5\%                   & 18.32                          & 1                                 & 17.53                            & 1                                   & 19.26                            & 1                                   \\ \cline{1-1}
LED 3000 k ULOR 5\%            & 18.15                          & 1.17                       & 17.49                            & 1.04                        & 18.22                            & 2.61                          \\ 
LED 4000 k ULOR 5\%            & 17.95                          & 1.41                      & 17.46                            & 1.07                        & 18.01                            & 3.16                        \\ 
LED 3000 k ULOR 0\%            & 18.39                          & 0.94                      & 17.73                           & 0.83                        & 18.45                            & 2.11                        \\ 
LED 4000 k ULOR 0\%            & 18.18                          & 1.14                      & 17.71                             & 0.86                        & 18.25                            & 2.54                        \\ \cline{1-1}
LED 3000 k ULOR 0\% L. Level 60\% & 18.95                          & 0.56                      & 18.25                            & 0.52                        & 18.98                            & 1.29                        \\ 
LED 4000 k ULOR 0\% L. Level 60\% & 18.71                          & 0.69                      & 18.23                            & 0.52                        & 18.78                            & 1.56                        \\ 
LED 3000 k ULOR 0\% L. Level 42\% & 19.41                          & 0.37                      & 18.74                            & 0.33                        & 19.43                            & 0.86                        \\
LED 4000 k ULOR 0\% L. Level 42\% & 19.22                          & 0.44                      & 18.74                            & 0.33                        & 19.27                            & 0.99                        \\ \cline{1-1}
\end{tabular}
\end{minipage}
\end{table*}

\section{Data bias from the broadband sensitivity of the SQM}

The spectral composition of night sky light varies with the meteorological conditions, and typically undergoes large changes during the passage of a front, but it can also appear unstable in a developing air mass or even in calm air due to fluctuations in the local aerosol system. Turbulent mixing, rising humidity, a change in the direction of the wind, the surrounding terrain, as well as the origin of the air mass have a direct impact on the actual optical properties of the aerosol \citep{bukowiecki2002real}, which, in turn, modifies the light field near the ground. Light emissions from ground-based sources predetermine the spectral properties of a diffuse light. However, the aerosols, dust particles, water droplets, and other atmospheric constituents change this picture, depending on the weighted contributions of the individual atmospheric components. For instance, maritime aerosol is dominated by water or sea salt both being slightly absorbing or non-absorbing materials, while urban aerosol typically originates from anthropogenic sources and thus may contain carbonaceous species. Therefore, maritime aerosols can act as efficient scatterers and thus intensify the night sky radiance. In contrast, urban aerosols may cause an intensity decay due to enhanced absorption of upward light emissions. The total effect of aerosol particles is often parametrized through the {\AA}ngstr\"om exponent, $\alpha$ (see, e.g. \cite{esposito2001preliminary}), which is used to model the spectral behavior of the aerosol optical depth (AOD).

\begin{equation}\label{eq:MK1}
AOD(\lambda)=AOD(500) \left( \frac{\lambda}{500} \right)^{- \alpha}		
\end{equation}
The wavelength $\lambda$ introduced above is measured in nanometers. \citet{cachorro2000measurements} have shown that $\alpha$ can range from -0.5 to more than 2, while Rayleigh theory dictates that the optical depth of the gaseous constituents (excluding absorption) is proportional to $\lambda^{-4.09}$ \citep{frohlich1980new}. To make things even more complex, a cloud layer has rather a neutral effect on the initial spectra because the water droplets dispersed in the clouds are large enough to scatter in the geometrical optics regime.

The above effects combine with the average spectral sensitivity of the SQM device, $T(\lambda)$, thus the signal recorded at the ground is

\begin{equation}\label{eq:MK2}
S_{SQM}=\int_{\lambda 1}^{\lambda 2} T(\lambda)\; \phi_0(\lambda)\; d \lambda
\end{equation}
where $\phi_{0}$ is the initial spectral signal (e.g. the sky radiance), and the integration interval $\lambda_{1}$ $\rightarrow$ $\lambda_{2}$ covers the SQM operating wavelengths. It has been shown by \cite{cinzano2005night} that $T(\lambda)$  varies with the angle of incidence of the beam, $\theta$, and thus a general functional form for $T(\theta,\lambda)$ should be considered in data processing, except for, e.g. SQM-L, that has a small field of view \citep{cinzano2007report}. As a consequence, the error margin of the approximation $T(\theta,\lambda) \rightarrow T(\lambda)$ is expected to be low for an SQM-L device.

\subsection{Considered lamp spectral distributions on this research}

The Lamp Spectral Power Distribution Database (LSPDD) is available online (www.lspdd.com) and is maintained by the Light Pollution Group of C\'egep de Sherbrooke. This dataset aims to provide independent information about the spectral characteristics of commercial lamp products. Each lamp has its associated datasheet providing the lamp's characteristics, like the spectral power distribution (SPD), correlated color temperature, percentage of blue light, and the impact indices introduced in \cite{aube2013evaluating}. The LSPDD distributes SPDs in ASCII text format (273 nm to 900 nm every 0.5 nm), allowing any other researcher to easily use the data for their own research. The LSPDD is released under the Creative Commons BY-NC-ND license. The data file is available in each lamp datasheet. All the SPD data are measured with the same spectrometer model (the Black Comet model from Stellarnet Inc.) factory calibrated according to NIST standards.    

\begin{figure}
 \includegraphics[width=250px]{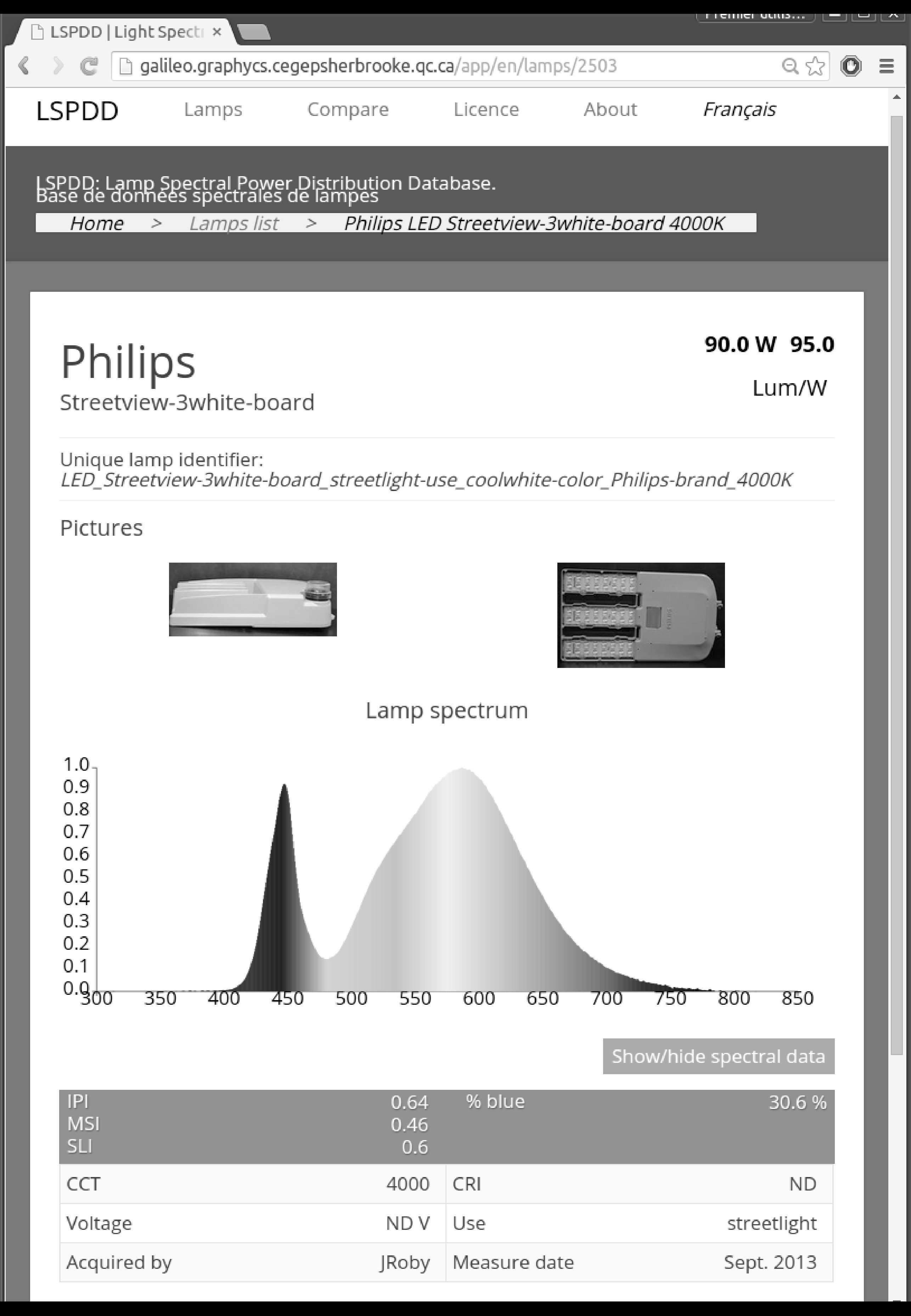}
 \caption{Screenshot of the web interface of the LSPDD. This window shows an example off a Philips LED 4000 K street light.}
 \label{lspdd}
\end{figure}

An example of the SPD for a white LED with $CCT \simeq 4000$ K from the LSPDD dataset is shown in Fig.~\ref{4000k} and a screenshot of the web interface is shown in Fig.~\ref{lspdd}.

The LSPDD is mainly based on lab tests and on new lamps. But, many of the lamps change their spectra with time. So, we have also used the LICA-UCM \citep{tapia2015lica} because it is based on street measurements. All the spectra used are normaliced to equal luminance on the pavement surface.

\subsection{Effect of SQM spectral sensitivity on cloud brightening factor measurements}

It has been pointed out in previous sections that the spectral sensitivity of a SQM device combines with atmospheric optical effects, while both form an output signal that is consequently used to characterize the diffuse light of a night sky. A synergy effect is a distortion of the original spectra, which makes the data interpretation partly uncertain. The light beams emitted from distant parts of an urban area propagate over longer optical paths and are attenuated (and scattered) more efficiently than those emitted from ilumination elements surrounding the measuring point. The zenith spectral radiance can therefore change with position even if the measurements are made within the territory of a city. The same is true for overcast skies. The brightening factor may vary with time and position, and of course with the wavelength \citep{aube2015clouds}.

\begin{figure}
 \includegraphics[width=250px]{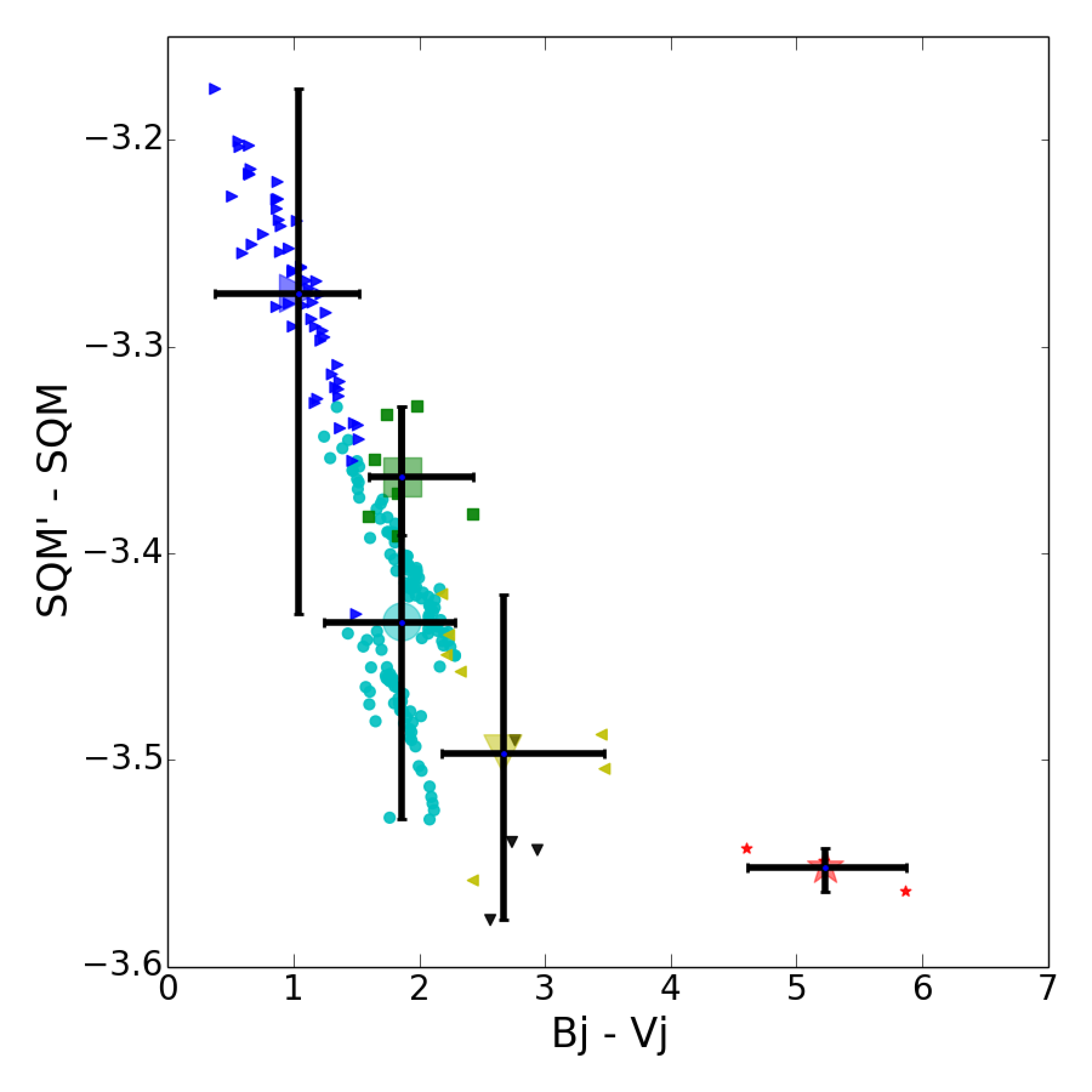}
 \caption{Change of the cloud brightening factor on synthetic photometry (see section \ref{SQM_char}) where SQM' represents the SQM reading with clouds, and SQM without clouds. This differential amplification is compared with the colour index $B_j-V_j$ in order to distinguish the different lighting technologies. The error bars represent the maximum range of possible values in each lamp category according to the SPD data available. This plot is for stratus clouds with cloud base height of 0.5 km, street lights with 5 per cent ULOR, and an AOD of 0.1. }
 \label{amplification1}
\end{figure}

In order to simulate this effect, we can use synthetic photometry, explained in Section \ref{SQM_char}, and we only need to multiply our source spectrum by the amplification monochromatic curve of \citep{aube2015clouds}. A typical result is shown in Fig.~\ref{amplification1}, in which $SQM_{CB}$ is the SQM reading with the SQM under a cloudy sky and the original measurements without clouds are denoted by SQM. 
 
\begin{figure}
 \includegraphics[width=250px]{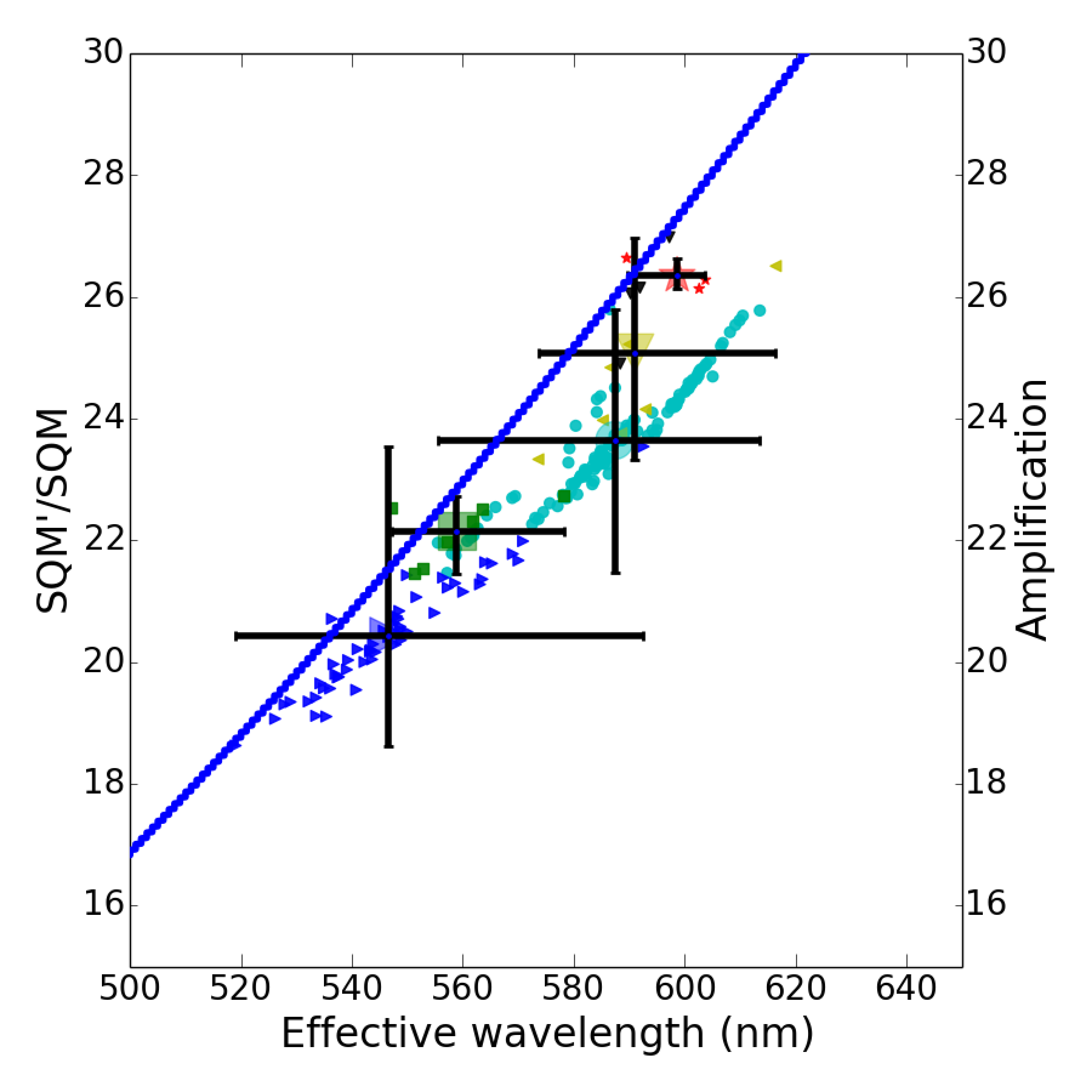}
 \caption{Change in amplification by the clouds on synthetic photometry (see section \ref{SQM_char}) where SQM’ represents the SQM reading with clouds, and SQM without clouds. Same plot as \ref{amplification1} but for comparison, the different technologies use the effective wavelength of each spectrum \citep{king1952note}.  The error bars represent the maximum range of possible values in each lamp category according to the SPD data available. The blue line represent the monochromatic amplification by the clouds; to interpret this curve, use the right axis. The amplification factors are for stratus clouds with cloud base height of 0.5 km, street lights with 5 per cent ULOR, and an AOD of 0.1.}
 \label{amplification2}
\end{figure}

An SQM reading sensitivity drops quickly for wavelengths above 600 nm, thus the sky appears artificially dimmer for these wavelengths. The effects of reddening are difficult to identify because the spectral response of SQM is higher at short wavelengths. This is why the measured amplification factors (AF) tend to be lower than those numerically computed for the whole visible spectrum (see Fig.~\ref{amplification2}). Namely, the theoretical values of AF steeply increase from blue to red, but the normalized response of SQM is far below unity at wavelengths above 600 nm. 

Fig.~\ref{amplification2} shows how the cloud AF vary with the lighting technology and, in turn, with the colour of the lamps. This plot is for the specific case of a stratus cloud with cloud base height of 500 m and a low aerosol loading characterized by an AOD of 0.1 in all cases. The lamp upward light output ratio (ULOR) was set to 5 per cent. The line is the theoretical monochromatic amplification calculated by \cite{aube2015clouds}. The wavelength associated to each lamp (points in the figure) correspond to the average emission wavelength of the lamp. The average emission wavelength is calculated using Eq.~\ref{averagew}, where $T(\lambda)$ must be replaced by ${\phi}(\lambda)$, the emission spectrum of the lamp. This plot shows that the SQM amplification factor is higher for redder lamps. The large point with error bars corresponds to the maximum emission wavelength of the averaged spectrum of all lamp in a given lamp group or technology. In summary, this plot shows that the observed amplification factor when using an SQM is underestimated by about 10 per cent for most of the lamps, but this percentage depends strongly on the specific lamp used. This effect is produced mainly for the non-monocromatic emission of these lamps.

\subsection{Effect of natural sources on SQM readings}

In this section we evaluate the impact of having different natural sources entering the Field of View (FOV) of the SQM. The impact of each source is driven both by the angular sensitivity curve and by the spectral response. In this paper we choose to focus mainly on the spectral response effect.

\subsubsection{Effect of the Field of View angular sensitivity}

Another source of error may be the natural sources of light present in the SQM field of view. By way of example we have made an analysis of the effect of the moon on the measurements of a SQM.

We characterized the SQM FOV used for this study as shown in Fig.~\ref{sqmangularresponse1}. The sensitivity drops very rapidly with the angle but if a very bright source lies at a large angle, it can contribute significantly to the total signal measured, even if the sensitivity is very low at that angle. As an example, we will estimate the impact of the full Moon at 50$^o$ of elevation, i.e. at 40$^o$ from the center of the FOV. The sensitivity at that angle is $\sim1$ per cent of the central value. Assuming a Moon magnitude of -12.74 \citep{williams2010moon}, when attenuated to 1 per cent (+5 magnitude) it gives an equivalent magnitude of -7.74. In other words, having the Moon at 40$^o$ produce the same signal to the SQM as having an object of mag -7.74 at zenith. In highly light polluted cities like Madrid, the SQM zenith sky brightness in very clear sky is around 18.5 mag~arcsec$^{-2}$ \citep{sanchezdemiguel2015variacion} while in the most pristine skies it is about 21.9 mag~arcsec$^{-2}$ \citep{benn1998palma}. To get a crude estimate of the comparative order of magnitude between the Moon at 40$^o$ and the sky brightness, let us determine what is the magnitude of the equivalent point source at zenith that would yield the same signal that the SQM reads from the sky.  We can calculate the equivalent magnitude of the sky from its surface brightness (mag~arcsec$^{-2}$) with the equation below: 

\begin{equation}
m_{SQM}=SB-2.5 \log \left(\frac{\pi D^2}{4} \right)
\label{MAE1}
\end{equation}
\noindent where $SB$ is the surface brightness in mag~arcsec$^{-2}$ and $D$ is the angular FOV of the SQM (FWHM$= 20^o$) in arcsec.

For the Madrid sky, we obtain a magnitude of -5.5 while for a pristine sky the equivalent magnitude is -2.1. This indicates that the difference in magnitude between the direct contribution of the Moon at 50$^o$ altitude and the contribution of the moonless sky is $\Delta m=-2.24$ (5.5--7.74) for Madrid and  $\Delta m=-5.64$ (2.1--7.74) for a pristine sky. That means that the contribution of the Moon to the SQM radiance at that specific altitude is $\sim8\times$ larger than the moonless sky brightness in Madrid and $\sim175\times$ larger in pristine skies. Note that this estimate does not consider the contribution of the moonlight scattered by the atmosphere into the FOV. 

\cite{pun2014contributions} used a model to evaluate the moonlight scattered to the zenith for many Moon altitudes and Moon phases for a given $V$ band extinction coefficient of $k = 0.58$\ mag/airmass. In particular, their fig. 7 shows that at 50$^o$ of altitude, the contribution of the zenith light scattered from the Moon reduces the sky brightness of 21.2 mag~arcsec$^{-2}$ (site Astropark Hong Kong) by $\sim3.5$ mag~arcsec$^{-2}$. If we apply Eq.~\ref{MAE1} to this sky brightness, we obtain an equivalent magnitude of -2.82, which leads to a magnitude difference with the equivalent magnitude of the direct light of the Moon attenuated for the angle of 40$^o$ from the FOV centre of $\Delta m=2.82-7.74=-4.92$. The  difference in magnitude between the direct moonlight at 40$^o$ of the centre FOV and the scattered moonlight is $\Delta m=3.5-4.92=-1.42$\ mag (i.e. a factor of $3.7\times$ on a radiance linear scale).

\subsubsection{Effect of the SQM spectral response}

The spectral impact of the different natural contributions to the zenith SQM readings requires access to a measured spectral database of the night sky radiance. We extracted this information from our archives; it was acquired with the SAND-4 spectrometers \citep{aube2007light,aube2015sand} for two extreme sites: an urban site located at UCM, and a rural site located in the Montsec astronomical park in Catalu\~na, Spain (42$^o$ 01' 28'' N, 00$^o$ 44' 10'' E). We combined information from the different sources present or absent from different spectra (e.g. with or without the Moon). Actually we used the full Moon night of June 3 to 4, 2015, and the new Moon night of June 16 to 17, 2015. For both sites, we used the data taken around the same time of the night ($\sim$ 23h50) to remove the effect of the variation of the sky brightness along the night. For the first night, the altitude of the Moon was 27$^o$ and the illumination of the Moon 98 per cent, whereas it was 0 per cent for the second night. Among other operations, the SAND-4 analysis software removes the continuous spectrum as its last step. This is done by fitting a low order polynomial on a set of wavelengths that do not show any contamination from the HID lamps, natural airglow, or Fraunhofer lines.

\begin{figure}
 \includegraphics[width=240px]{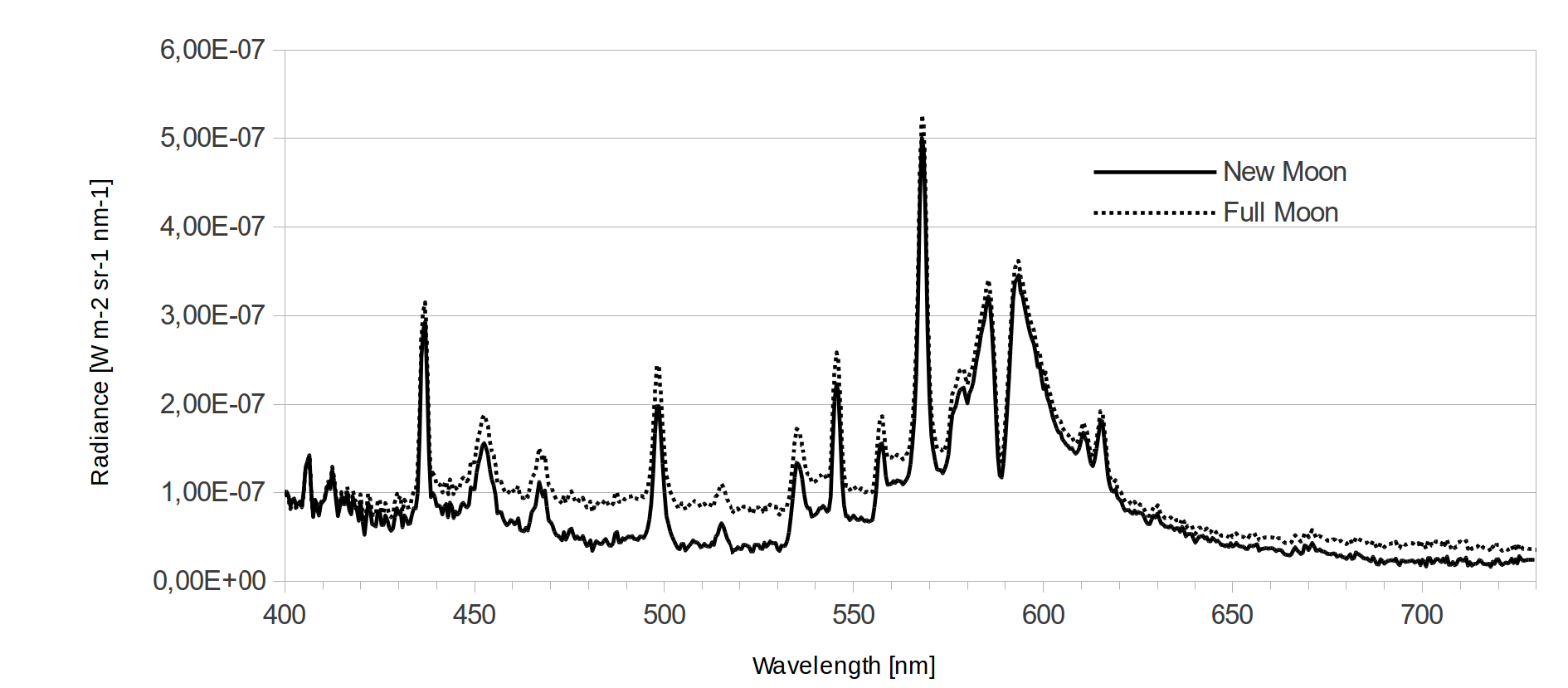}
 \caption{Observed zenith radiance in UCM during New Moon (2015 June 3) and Full Moon (2015 June 16) around 23h50 UTC. }
  \label{observedUCM}
\end{figure}

\begin{figure}
 \includegraphics[width=240px]{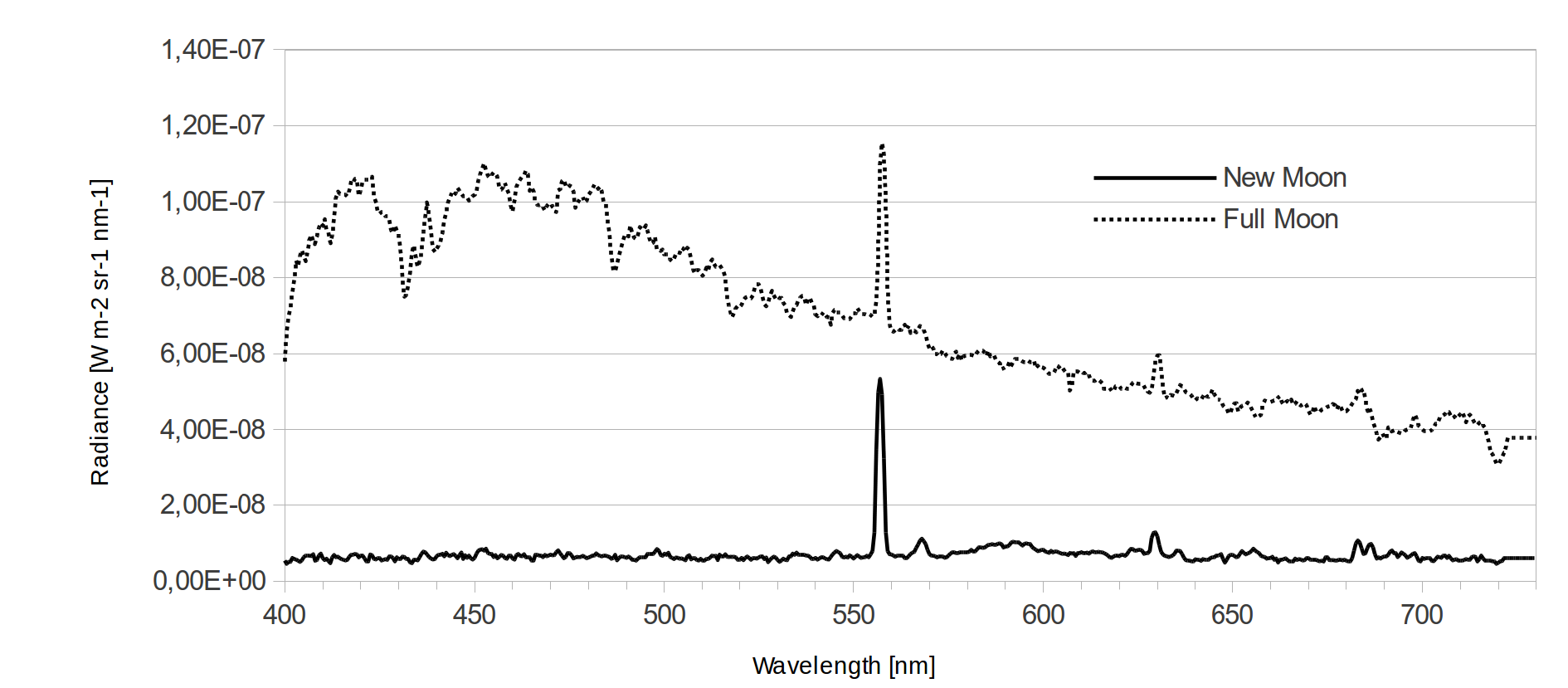}
 \caption{Observed zenith radiance in Montsec during New Moon (2015 June 3) and Full Moon (2015 June 16) around 23h50 UTC. }
 \label{observedmontsec}
\end{figure}

We used a multi-Gaussian fitting algorithm to approximate the line spectrum as a combination of many Gaussian lines (actually 30 Gaussian profiles). In that process, we locked the FWHM of the lines to a constant value, so that this parameter is determined by the optics of SAND-4. The retained FWHM for the UCM instrument was 1.5 nm while it was 1.25 nm for the Montsec instrument. With a detailed knowledge of the contribution of the spectral lines to the spectrum, it was then possible to evaluate separately the contribution of the lines associated with human activities and the natural airglow. The artificial and natural contributions to the continuous spectrum (e.g. car headlights, halogen and incandescent, LEDs, star background) was obtained by subtracting the spectral lines fitted from the new Moon observations. To estimate the typical star background contribution, we assumed that the continuous spectrum of the new Moon in Montsec is dominated by the star background. To estimate the artificial part of the continuous spectrum, we started from the new Moon measurement in Madrid and subtracted the spectral lines fitted along with the star background previously found with the Montsec measurement. Finally the artificial portion of the continuous spectrum in Montsec was estimated by re-scaling the artificial continuous spectrum found for Madrid using the averaged ratio of the artificial line spectra for the two sites. By doing this, we then assumed that the proportion of the artificial line radiance to the artificial continuous radiance is more or less the same for both sites. Lastly, to determine the contribution of the scattered moonlight, we subtracted the new Moon spectrum from the full Moon spectrum. Note that we took care to acquire the spectrum at the same time of night for both nights, to exclude the effect of the decrease in light levels along the night. At the end of that process we had the following basic spectra. 1: the artificial line and artificial continuous spectrum at UCM, 2: the artificial line and artificial continuous spectrum at Montsec, 3: the natural sky glow, 4: the star background, and 5: the full moon scattered moonlight at 27$^o$ of Moon altitude.

\begin{table*}
 \centering
 \begin{minipage}{140mm}
  \caption{Impact of natural sources on the SQM measurement in comparison to a pure artificial sky brightness under clear sky conditions}
  \begin{tabular}{@{}llrrrrlrlr@{}}
  \hline
  Site	&  	Madrid	&&  Montsec	& \\
 \hline 
  & 	$\Delta SB$ 	& added to & 	$\Delta SB$ 	& added to  \\

   								& 	& artificial & 	 	& artificial \\
     							& 	mag~arcsec$^{-2}$ 	& \% & 	mag~arcsec$^{-2}$ 	& \%  \\             
 \hline
  Artificial + Natural 			& 	+0.007 	& 0.7 & +0.43 	& 49	\\
  Artificial + Natural + Stars	& 	+0.07 	&  7 & 	+1.90 	& 475	\\
  Artificial + Natural + Stars + full moon Scat. Light & 	+0.66 &  83 & 	+4.42 	& 5760	\\
\hline
\end{tabular}
\label{natural-effect}
\end{minipage}
\end{table*}

\begin{figure}
 \includegraphics[width=240px]{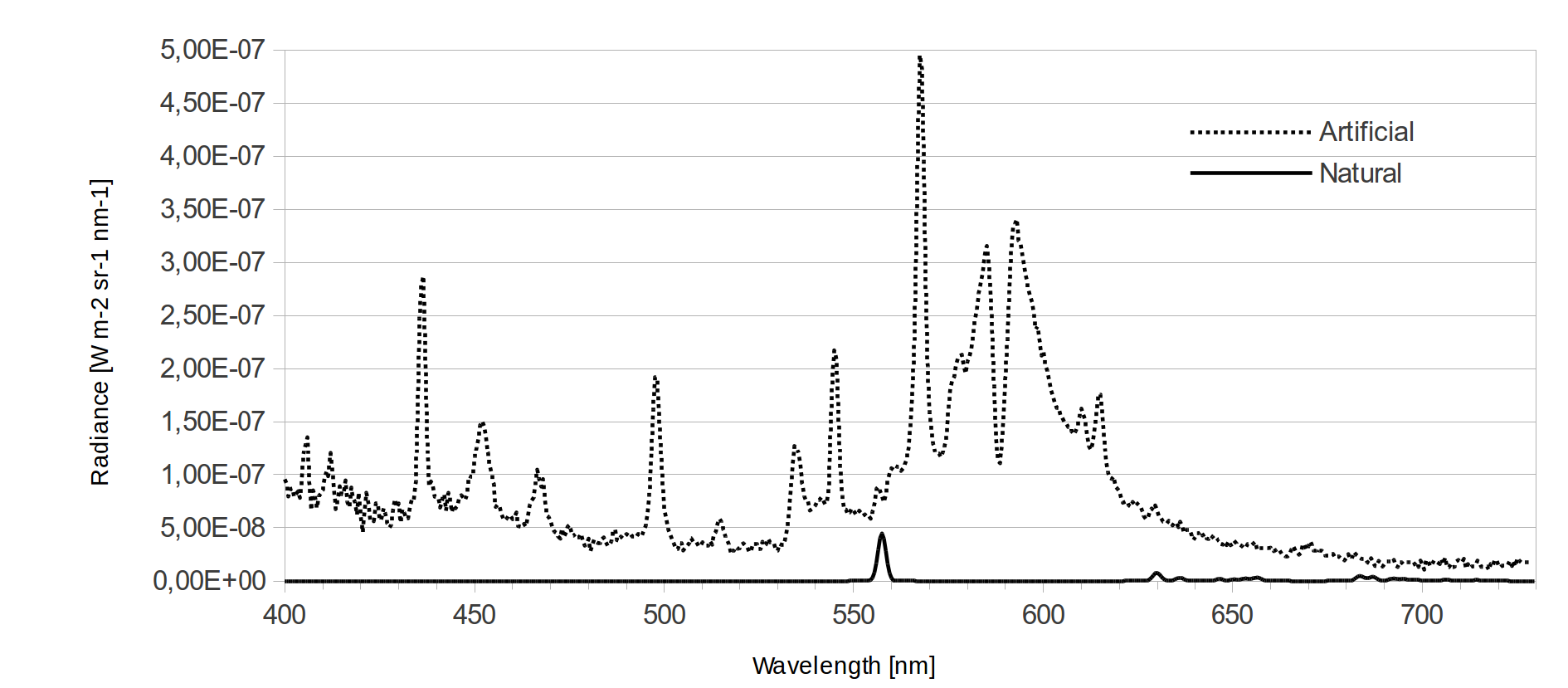}
 \caption{Zenith radiance from artificial and natural airglow in UCM extracted from Figs \ref{observedUCM} and \ref{observedmontsec}. }
 \label{artnatUCM}
\end{figure}

\begin{figure}
 \includegraphics[width=240px]{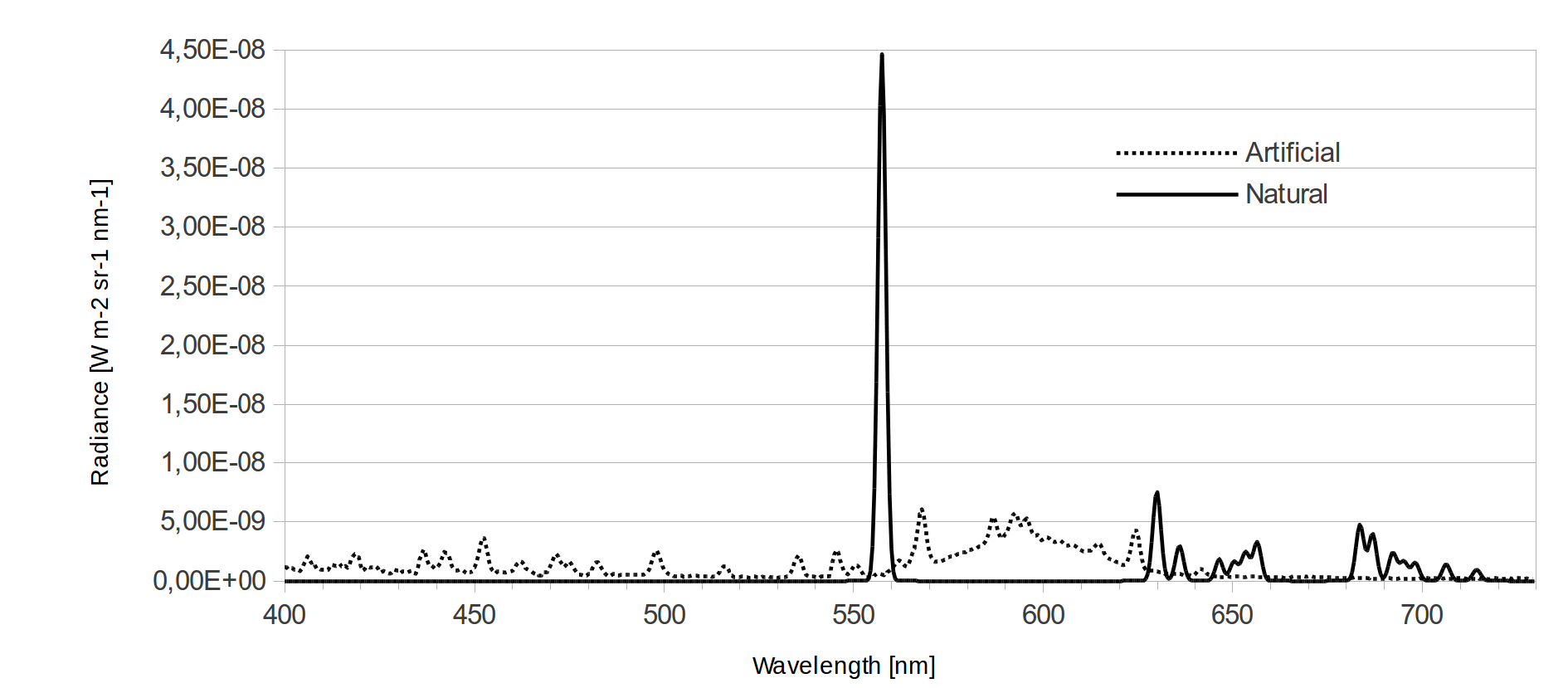}
 \caption{Zenith radiance from artificial and natural airglow in Montsec extracted from Figs \ref{observedUCM} and \ref{observedmontsec}. }
  \label{artnatmontsec}
\end{figure}

\begin{figure}
 \includegraphics[width=240px]{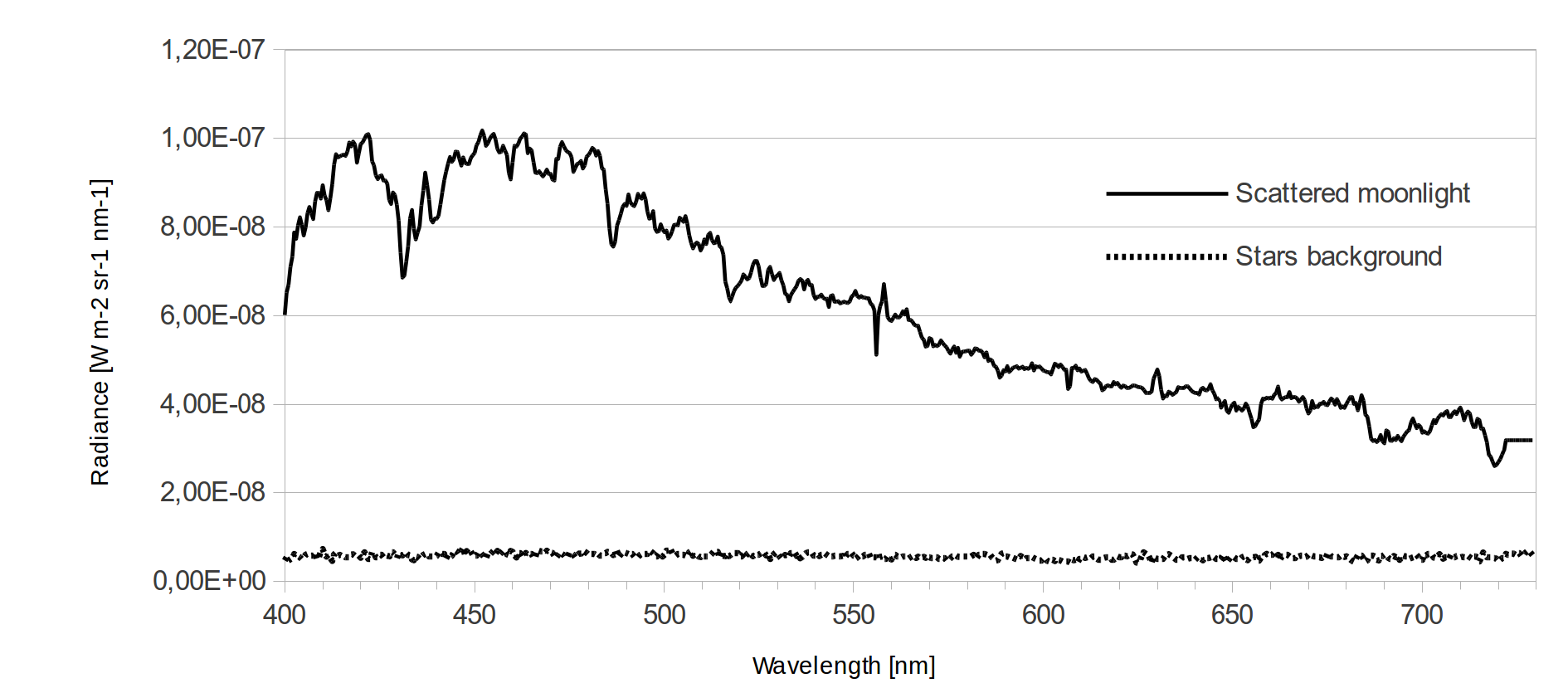}
 \caption{Scattered moonlight and star background zenith radiance extracted from Figs \ref{observedUCM} and \ref{observedmontsec}. }
 \label{moonstars}
\end{figure}

\begin{figure}
 \includegraphics[width=240px]{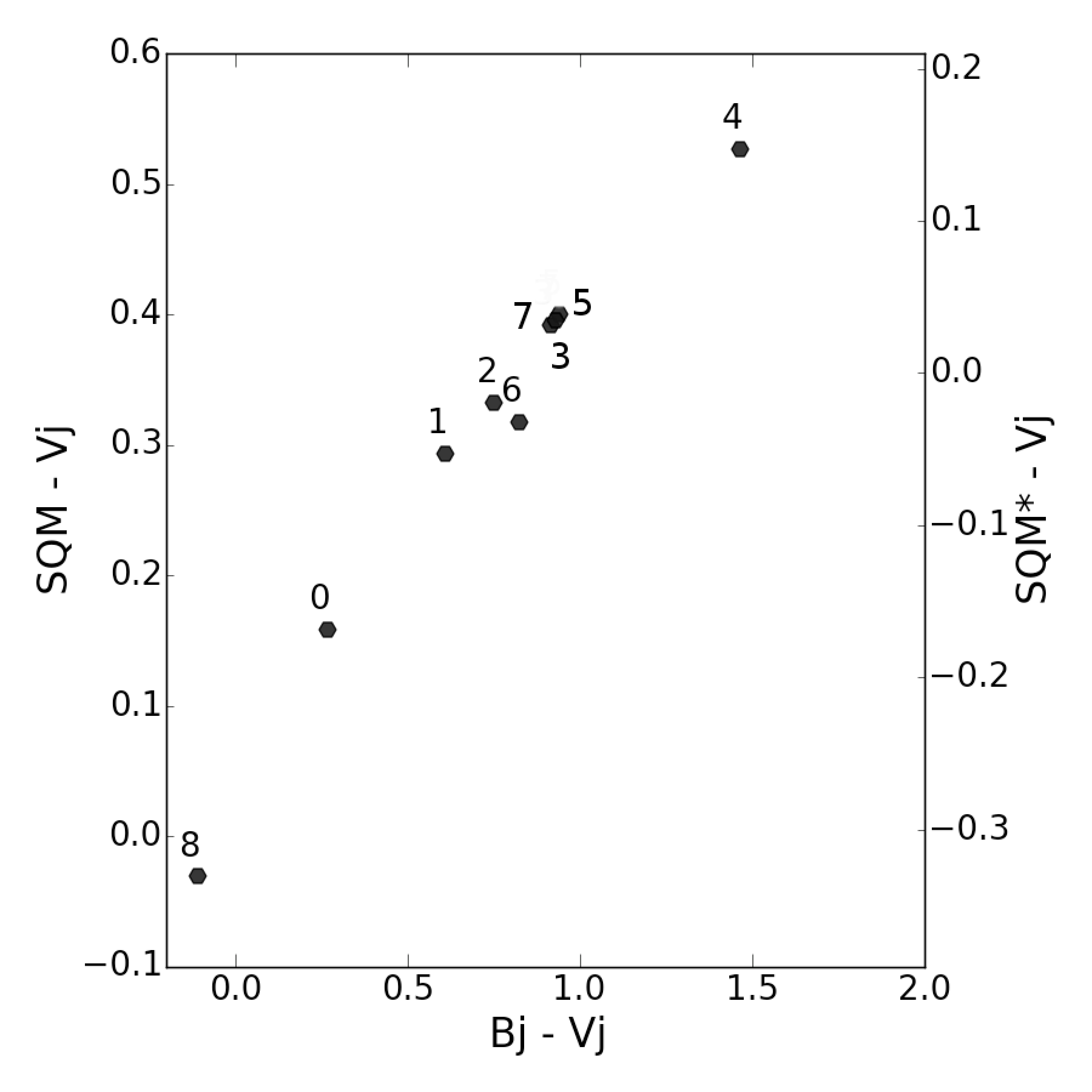}
 \caption{ ${SQM}-V_j$ colour vs $B_j-V_j$ colour for different sky radiance components, 
 where 0 = A+N+S+SM in Montsec, 1 is  $A + N + S + SM$ in UCM, 2 is $A + N + S$ in Montsec, 
 3 is $A + N + S$ in UCM, 4 is $A + N$ in Montsec, 5 is $A + N$ in UCM, 6 is $A$ in Montsec, 7 is $A$ in UCM, 
 8 = Vega spectrum. Legend: $A$ = artificial radiance, $N$ = Natural sky lines radiance, $S$ = Star background radiance, 
 $SM$ = Scattered moonlight }
 \label{moonstars2}
\end{figure}

The objective when measuring the sky brightness of a given site is in most cases to follow up with a determination of the artificial contribution to the sky brightness. But it is not obvious how to know what fraction of the SQM measurement is really of human origin. To get a good idea of this, we combine the basic spectra to mimic realistic situations and compare this to what the SQM 
should detect without any natural sky brightness (i.e. only the artificial brightness). The retained combinations are 1: the artificial and the natural airglow spectrum, 2: the artificial, natural airglow and the star background, and 3: the artificial, natural airglow, 
the star background and the full moon scattered moonlight at $27^o$ of Moon altitude. Table~\ref{natural-effect} shows the shift in sky brightness 
($mag_{SQM}$~arcsec$^{-2}$) of the retained combination in relation to the artificial SB alone. To retrieve the artificial SB from the measured SQM SB you must add $\Delta SB$ to the SQM measurement. 
The table also indicates the percentage increase in the observed radiance compared to the artificial radiance for the given source combination and site. 
If one want to determine the percentage of a specific natural source it is possible to substract one ``added to artificial \%'' from the previous line in the table. 
As an example, if one want to know the percentage of the contribution from stars, it is given, in the case of Madrid, by $7\%-0.7\%=6.3\%$. 
We can see that for an urban site like Madrid, the natural airglow contribution to the SQM radiance is very weak (less than 1 per cent). 
Even the stars' background contribution is relatively small (less than 10 per cent) but, as expected, the full moon contribution is more important, 
actually almost as much as the artificial radiance itself ($83\%-7\%-0.7\%\simeq 75\%$). The results are completely different for a dark site like Montsec. 
In such a case, the natural airglow is approximately one-half of the radiance of artificial origin. The star background is in that case approximately $4 \times$ the artificial 
radiance ($475\%-49\% \simeq 400\%$). Finally it is not surprising to see that the scattered radiance of the full Moon is very high compared to the artificial 
radiance ($\sim 50 \times$, i.e. $5760\%-475\%-49\% \simeq 5200\%$).

\section{Discussion}

SQM photometers have become the reference instrument for measuring the night sky brightness, and networks of fixed stations are being built to monitor the evolution of the sky brightness. Due to the single broad band(monochannel) of the SQM, the eventual lower SQM readings will not necessarily translate to that the sources of light pollution really reduce light pollution.  

Our cities are experiencing a change in their artificial illumination, both in intensity and technology. This results in a different colour of the night sky or, more precisely, to different night sky spectra.  We have shown that the usual photometric method for correcting from colour using colour terms or magnitude offsets according to the type of lamps is imprecise due to the large scatter in colour among each lamp type.

To be on the safe side, we recommend using error bars based on the lighting technology instead of a colour corrected version when the spectra, intensities, and proportions of the light pollution sources are unknown.

It is interesting to note that the massive replacement of street illumination with white LEDs will result in darker SQM readings and for the photopic vision due to the important blue component in their spectra. However the resulting sky is brighter for the human eye scotopic vision for the same radiance.

Does this mean that the SQM should not be used? As long as an improved device is not yet available, we should continue with the SQM. But, we need to train ourselves to avoid interpretation errors of the data and use it only as complementary data to colour sensitive sensors. Another option is to start using astronomical filters as has already been done by \cite{kyba2012red}.

The SQM is the first device that has been able to measure massive spread and it continues to be a valuable instrument for tracing variations as long as the spectra of the sources do not undergo extreme changes. Nonetheless, it has become clear that the colour of light should be regarded as a key parameter in light pollution studies. Therefore, the use of colour sensitive devices should be promoted, especially, in places where the light spectra are undergoing changes \citep{spoelstra2014new,zamorano2015low}. The data already acquired with SQMs are scientifically valid as is shown in  \cite {falchi2016new}. But, as is explained in this article and also in  \cite {falchi2016new} the lack of colour information with the use of satellites and the SQM might be hiding and increase of light pollution. An increase that, in extreame cases, can be as high as 261\%.

\section{Conclusions}
Monitoring light pollution levels has become important in the last decades because of rapid industrialization and modernization, especially in densely populated regions. Undoubtedly, the light emissions upwards and the spatial extent of light pollution are both increasing functions of the amount of light emitted and the number of sources.  However, an overillumination of the environment is also somehow correlated with population growth, thus the optical methods of detection are a desirable technology for watching the effects caused by atmosphere and/or ground-based light sources. Optical methods are preferred because such technology is rapid, the components are inexpensive, and the techniques can be automated. 

The Sky Quality Meter (SQM) is a widely available device that works properly in the low-intensity regime. Although SQM is cheap enough and operated worldwide, the data are far from being error free. Specifically, the SQM has a fast readout and an acceptable dynamic range, but low angular and spectral resolutions. This has a direct impact on the interpretation and comparison of light pollution levels measured under distinct conditions. Note that lighting technologies and lamp installations may deviate significantly from case to case. As a result, the same SQM signals might not necessarily mean the same levels of light pollution, because the former can originate from arbitrarily weighted contributions of different lamp types. In addition, atmospheric optics make things more difficult, mainly due to the non-trivial distortion of optical signals when traveling along different trajectories. In addressing this task, we have become aware of how little we know of the underlying physics and variability of SQM signals.

This paper addresses directly this shortcoming and is aimed at a correct understanding of the observed data. It was shown that the blue light emitted abundantly from artificial sources makes the sky colour brighter thanks to low-intensity light vision (scotopic vision), while the source itself appears darker because it is subject to photopic vision. The latter is also applicable to the $V$-band. We have also shown that SQM readings vary with the colour of the light pollution sources, but these data can scarcely be predicted as a simple average over the technologies applied. Atmospheric effects impose an additional optical distortion that can change the sky colours and even SQM readings. Clouds are recognized as having the greatest influence on the ground reaching signals, while diffusely reflected light more-or-less mimics the spectrum of artificial sources. This is why the brightering factor (BF) alters with the spectral features of artificial lamps that, in turn, are important sources of uncertainty in SQM readings. This imposes an additional burden on the BF as determined from SQM data, because different effects are involved in forming the zenith radiance under cloudy and cloud-free conditions. There is no doubt that a colour-sensitive device is needed in order to reduce the posible miss interpretation that otherwise arise from SQM data. The SQM is mesuring the sky brightness for the SQM band correcly, but it is not representative of the human vision with enouth acuracy when the spectra os the sources is more relevant.  Such a device can make the sky brightness retrieval more accurate and thus more suitable for monitoring the optical effects by different installations of ground-based light sources. Although the single measurements of programs like Globe at Night can be very imprecise, 
the amateur measurements are still fundamental to trace the real long term change of the sky quality as the human eye physiology is not expected to change in thousands of years \citep{kyba2013citizen,falchi2016new}.

As final conclusion, the SQM is a good instrument but is not good enough to trace the evolution of a change of sky brightness in a color changing world. 
Color sensors or multiband cameras are needed to trace the real evolution of the sky brightness in this color changing world. 
However more precise human based mesurements of the sky brightness, like the Naked Eye Limiting Magnitude IMO's tecnique \citep{rendtel1995handbook} and/or Loss of the Night tecnique \citep{loss22}, 
is still fundamental to trace long term evolution of the night sky quality.

\section*{Acknowledgments}
This research was supported by Fonds de recherche du Qu\'ebec -- Nature et technologies (FRQNT) and by the F\'ed\'eration 
des C\'egeps.png, by the Slovak Research and Development Agency under contract No. APVV-14-0017, by Spanish projects AYA2012-31277 
and AYA2013-46724-P and the Spanish Network for Light Pollution Studies (AYA2015-71542-REDT) from the 
Spanish Ministerio de Econom\'ia y Competitividad. This research also received support from an MVTS grant of the Slovak Academy of Sciences. 
This research was supported in part by the STARS4ALL project and ORISON project funded by the European Union H2020-ICT-2015-688135 and H2020-INFRASUPP-2015-2.

We thank A. Morin-Paulhus, a former student from the C\'egep de Sherbrooke,
who developed the web-based interface of the LSPDD database, along with
S. Ribas for the maintenance of the SAND instrument in Montsec. We also thank Emma Rebecca Howard and the referee for their comments.
\nocite{*}
\bibliographystyle{mn2e}
\bibliography{references_clean-R2}

\label{lastpage}

\end{document}